\documentclass[aps,twocolumn,groupedaddress,floatfix,showpacs,showkeys,amsmath,amssymb,prb]{revtex4-1}
\usepackage{amsmath}
\usepackage{amssymb}
\usepackage{graphicx}
\usepackage{textcomp}
\usepackage{bm} 
\usepackage{braket} 
\usepackage[usenames,dvipsnames]{color}

\begin{document}

    \title{Statistics of Anderson-localized modes in  disordered  photonic crystal slab waveguides}

    \author{J. P. Vasco}
    \email{jpvc@queensu.ca}
\affiliation{Department of Physics, Queen's University, Kingston, Ontario, Canada, K7L 3N6}
\author{S. Hughes}
\email{shughes@queensu.ca}    
\affiliation{Department of Physics, Queen's University, Kingston, Ontario, Canada, K7L 3N6}

\pacs{xxxx, xxxx, xxxx}                                         
    
    \pacs{42.70.Qs, 42.25.Fx, 42.82.Et, 42.81.Dp}
\keywords{photonic crystal waveguides; scattering; disorder; Anderson localization; quantum dots }
\begin{abstract} 
We present a fully three-dimensional Bloch mode expansion technique and  photon Green function formalism to compute the quality factor, mode volume, and Purcell enhancement distributions of a disordered W1 photonic crystal slab waveguide in the slow-light Anderson localization regime. By considering fabrication (intrinsic) and intentional (extrinsic) disorder we find that the quality factor and Purcell enhancement statistics are well described by log-normal distributions without any fitting parameters. We also compare directly the effects of hole size fluctuations as well as fluctuations in the hole position. The functional dependence of the mean and standard deviation of the quality factor and Purcell enhancement distributions is found to decrease exponentially on the square root of the extrinsic disorder parameter. The strong coupling probability between a single quantum dot and an Anderson-localized mode is numerically computed and found to exponential decrease with the squared extrinsic disorder parameter, where low disordered systems give rise to larger probabilities when state-of-art quantum dots are considered. The optimal regions to position  quantum dots in the W1 waveguide are also discussed. These theoretical results are fundamental interesting and connect to recent experimental works on photonic crystal slab waveguides in the slow-light regime. 
\end{abstract}

\maketitle

\section{Introduction}

Periodic dielectric optical structures, known as photonic crystals (PC), have been focus of intense research over the last four decades since Yablonovitch \cite{yablonovitch} and John \cite{john} investigated their capabilities to profoundly  modify the local density of electromagnetic states (LDOS); these works  have served as inspiration to develop a detailed optical theory of PCs \cite{joannopoulos}. In particular, PC slabs, which are practical two-dimensional PCs embedded in semiconductor dielectric membranes, have become an excellent candidate for on-chip integration given their flexibility to tailor the confinement and propagation of light, aided by continuously improving   fabrication techniques \cite{reinhard}. A wide range of cavity quantum electrodynamics (cavity-QED) phenomena have been observed in PC slabs using quantum dots (QDs), in both weak and strong coupling regime \cite{yoshie,hennessy}, as have resonant interactions between several quantum emitters within PC slabs \cite{villasboas}, with promising functionalities in quantum information processing \cite{hughes2,jpvasco}. Rich optical non-linear phenomena have also been experimentally and theoretically addressed in these systems \cite{galli,flayac}. 

With regards to waveguide geometries, PC slab waveguides (PCWs) have  attracted much attention as a platform for single photon generation \cite{lodahl4,hughes3}, optical buffering \cite{baba}, and for enhancing  non-linear optical processes in the slow-light regime \cite{krauss,colman,monat} However, PCWs are known to be extremely sensitive to  fabrication disorder \cite{gerace,faolain,talneau}, which becomes relevant for practical applications \cite{hughes4,ramunno,lalanne1}. Interestingly, structural disorder in PCs can also induce coherent scattering of light \cite{patterson1}, giving rise to the  formation of random cavities where the field can be strongly localized, leading to the quasi-1D Anderson localization regime \cite{john,segev}. Currently, the Anderson localization phenomenon have been intensively studied in PCWs with potential applications on cavity-QED \cite{sapienza} and lasing \cite{liu}, especially near the photonic band-edge frequency region, where the density of photonic states (DOS) become very sensitive to structural disorder \cite{garcia}. Furthermore, the role of intentional or deliberate disorder has also been addressed in the literature \cite{nishan1} with promising results for forming high-$Q$ cavity modes \cite{topolancik1,topolancik2} (where $Q$ is the quality factor) and transverse localization below the diffraction limit \cite{hsieh}.

Despite the relevance of the various model studies on disordered PCWs, many of the numerically exact theoretical calculations are carried out by using  two-dimensional models \cite{Garcia1,lalanne2} or even one-dimensional ones \cite{lodahl1}. This is manly due to the high computational cost required to solve large photonic structures in three-dimensional calculations, e.g., 3D-FDTD calculations, and the statistical nature of the Anderson phenomenon, where several realizations of the disordered system must be calculated to obtain accurate results. Thus, the numerically easier (faster) two- and one-dimensional approaches allow one to carry out statistical analysis with moderate computational effort. Nevertheless, these approaches are limited, and they cannot  quantify  the out-of-plane losses, coming from coupling to modes above the   slab light line, and the actual electric field value throughout the photonic structure can also be quite different. These two important quantities and the frequency spectrum of the system determine the quality factor, mode volume and Purcell enhancement (PE) statistics, which, from our knowledge, have not been rigorously studied in three-dimensional models for large scale disordered PCWs. Yet, they are critically important in understanding the properties of such modes, especially for assessing how such modes couple to single quantum emitters such as  QDs. 

In the present paper, we present a fully three-dimensional Bloch mode expansion method and a Green function formalism to explicitly calculate the quality factor, mode volume and PE distributions of a  disordered W1 PCW; fabrication (intrinsic) and two models of intentional (extrinsic) disorder are considered in our calculations, including variations in hole radii as well as position. We find that the cavity-mode quality factor statistics is in general well described by log-normal distributions, as well as the PE statistics for moderate quantities of intentional disorder. The latter is a consequence of the dominant behaviour of $Q$ over the PE given the small variation range of the mode volumes in comparison to the variation range of the quality factors. In particular, we show that the intensity of the emission enhancement is slightly larger in the $\hat{x}$ projected direction (waveguide direction) than in the $\hat{y}$ projected one (perpendicular to the waveguide direction within the PC plane), which is due to the larger value of the $x$-polarized electric field component peak in comparison to the corresponding $y$ component. A decreasing exponential dependence of the mean quality factors and mean PE on the square root of the extrinsic disorder parameter is found, as well as for the standard deviation of the corresponding statistical distributions. Finally, using our three-dimensional model, we also compute the probability of strong coupling between state-of-art QDs and to the Anderson-localized modes, and find the optimal spatial regions to maximize coupling of these quantum emitters in the PCW. A decreasing exponential dependence of the strong coupling probability on the squared extrinsic disorder parameter is seen, and such probability is found to be more robust against hole-position disorder than hole-size disorder. Our theoretical results are of particular relevance to complement recent experimental works on probing the statistical properties of Anderson localization \cite{lodahl2}, statistical measurements of Purcell enhancements \cite{lodahl3} and cavity-QED \cite{sapienza}.

The rest of our paper is organized as follows. In Sec.~\ref{BME}, we briefly discuss the Bloch mode expansion method recently introduced by  Savona \cite{savona1,savona2}. In Sec.~\ref{SSEnt}, we  study the non-disordered W1 PCW using the guided mode expansion method, compute the Bloch mode basis to solve the disordered W1 PCW, and establish the intrinsic and extrinsic disorder models. The statistics of the quality factor and mode volumes of disorder-induced cavity modes are presented in Sec.~\ref{SQV}, and the disordered DOS, as well as the statistics of PEs are presented in Sec.~\ref{DOSPE}. Finally, we compute the probability of strong coupling between a quantum emitter and the Anderson-localized modes in Sec.~\ref{SC}. The main conclusions of the work are presented in Sec.~\ref{Concl}.

\section{The Bloch mode expansion method}\label{BME}

The eigenmodes of perfect (i.e., non-disordered) PCs can be described by the stationary wave equation \cite{joannopoulos}
\begin{equation}\label{maxwellperfect}
\nabla\times\left[\frac{1}{\epsilon(\mathbf{r})}\nabla\times\mathbf{H}_{\mathbf{k}n}(\mathbf{r})\right]-\frac{\omega^2_{\mathbf{k}n}}{c^2}\mathbf{H}_{\mathbf{k}n}(\mathbf{r})=0,
\end{equation}
subject to the Bloch boundary condition, i.e., $\mathbf{H}_{\mathbf{k}n}(\mathbf{r}+\mathbf{R})=e^{i\mathbf{k}\cdot\mathbf{R}}\mathbf{H}_{\mathbf{k}n}(\mathbf{r})$, where $\mathbf{R}$ is the lattice vector. Equation~(\ref{maxwellperfect}) applies for linear and isotropic non-magnetic materials, and we also assume that the dielectric function describing the system is real and positive (transparent materials) without any explicit dependence on $\omega$ (non-dispersive material). The eigenstate $\mathbf{H}_{\mathbf{k}n}$ is the magnetic Bloch mode with band index $n$ and momentum $\mathbf{k}$, and the corresponding eigenvalue defines the resonant frequency of the mode. The periodic profile is represented by $1/\epsilon(\mathbf{r})$, where the dielectric function $\epsilon(\mathbf{r})$ has the same periodicity of the non-disordered PC. In the presence of disorder, the new dielectric profile $1/\epsilon'(\mathbf{r})$ is written as
\begin{equation}\label{etadef}
\frac{1}{\epsilon'(\mathbf{r})}=\frac{1}{\epsilon(\mathbf{r})}+\eta(\mathbf{r}),
\end{equation}
and Eq.~(\ref{maxwellperfect}) can be applied with Bloch boundary conditions, in a disordered supercell with dielectric function $\epsilon'(\mathbf{r})$, for the disordered eigenstates $\mathbf{H}_{\beta}$, yielding
\begin{equation}\label{maxwelldis}
\nabla\times\left[\frac{1}{\epsilon'(\mathbf{r})}\nabla\times\mathbf{H}_{\beta}(\mathbf{r})\right]-\frac{\omega^2_{\beta}}{c^2}\mathbf{H}_{\beta}(\mathbf{r})=0.
\end{equation}
The eigenmodes in Eq.~(\ref{maxwelldis}) are labeled with a new index $\beta$ since $\mathbf{k}$ is not conserved when disorder is taken into account. Following Savona \cite{savona1,savona2}, the disordered modes can be expanded in an orthonormal basis of Bloch modes, using the solutions of Eq.~(\ref{maxwellperfect}):
\begin{equation}\label{bmeexpansion}
\mathbf{H}_{\beta}(\mathbf{r})=\sum_{\mathbf{k},n}U_{\beta}(\mathbf{k},n)\mathbf{H}_{\mathbf{k}n}(\mathbf{r}),
\end{equation}
turning Eq.~(\ref{maxwelldis}) into the following linear eigenvalue problem:
\begin{equation}\label{eigenBME}
\sum_{\mathbf{k},n}\left[V_{\mathbf{k}n,\mathbf{k}'n'}+\frac{\omega^2_{\mathbf{k}n}}{c^2}\delta_{\mathbf{k}\mathbf{k}',nn'}\right]U_{\beta}(\mathbf{k},n)=\frac{\omega^2_{\beta}}{c^2}U_{\beta}(\mathbf{k'},n'),
\end{equation}
with disordered matrix elements defined a
\begin{equation}\label{matrixV}
V_{\mathbf{k}n,\mathbf{k}'n'}=\int_{\rm s. cell}\eta(\mathbf{r})\left(\nabla\times\mathbf{H}_{\mathbf{k}n}(\mathbf{r})\right)\cdot\left(\nabla\times\mathbf{H}^{\ast}_{\mathbf{k}'n'}(\mathbf{r})
\right)d\mathbf{r},
\end{equation}
where the integration is carried out over the entire supercell.
In the same way as it is considered in Refs.~\onlinecite{savona1}~and~\onlinecite{minkov1}, we adopt the guided mode expansion (GME) approach to compute the integral of Eq.~(\ref{matrixV}), which conveniently turns the matrix elements into analytical closed expressions \cite{andreani}. Such an approach has shown to be very accurate and efficient to describe photonic eigenmodes in high-dielectric-contrast PC slabs with complex two-dimensional profiles \cite{agio,gerace,galli}, as  is the case of the photonic structures in the present work. Once the disordered magnetic field is computed, the electric field is obtained via Maxwell's equations, so that
\begin{equation}\label{efield}
\mathbf{E}_{\beta}(\mathbf{r})=\frac{ic}{\omega_\beta\epsilon'(\mathbf{r})}\nabla\times\mathbf{H}_{\beta}(\mathbf{r}),
\end{equation}
which is  normalized through
\begin{equation}\label{enorm}
\int_{\rm s. cell}\epsilon'(\mathbf{r})|\mathbf{E}_{\beta}(\mathbf{r})|^2d\mathbf{r}=1.
\end{equation}
This normalization allowing the calculation of the effective mode volume,  V$_\beta=1/(\epsilon'(\mathbf{r}_0)|\mathbf{E}_{\beta}(\mathbf{r}_0)|^2)$, where $\mathbf{r}_0$ is usually taken at  the antinode position of the electric field peak. The imaginary part of the eigenfrequency, associated to the assumed real eigenvalue in Eq.~(\ref{eigenBME}), leads to out-of-plane losses for PC slabs, and can be estimated using the same formulation of the photonic golden rule given in Ref.~\onlinecite{andreani}, where the radiative decay from a disordered mode to a radiative mode is calculated above the light line using first-order time dependent perturbation theory.
This approach is expected to be a good approximation when the real part of the eigenvalue is much larger than its imaginary part (low-loss regime). The complex frequencies of the system are then approximately defined as $\tilde{\omega}_\beta=\omega_\beta-i\Omega_\beta$, where the quality factor of the disordered photonic eigenmode is computed by $Q_\beta=\omega_\beta/(2\Omega_\beta)$. While such cavity modes are formally quasinormal modes \cite{kristensen1}, for high $Q$ resonators, the effective mode volume defined earlier is expected 
to be an excellent approximation for the typical spatial domains we use \cite{kristensen2}.

\section{Deliberately disordered W1 photonic crystal slab waveguides}\label{SSEnt}

\begin{figure}[t]
  \begin{center}
    \includegraphics[width=0.48\textwidth]{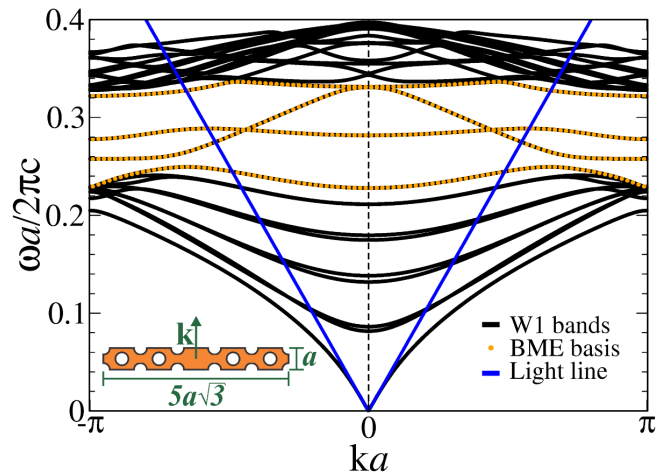}
  \end{center}
  \vspace{-0.3cm}
  \caption{
(Color online) (Black lines) Projected photonic dispersion of the one-period regular W1 PCW by considering a regular supercell of dimensions $a\times5a\sqrt{3}$, as it is shown in the inset. (Blue lines) Light lines of the slab. (Orange points) Set of modes considered in the BME calculation.}
  \label{fig:bandsW1}
\end{figure}

In order to build a suitable Bloch mode basis to describe a disordered air-bridge W1 system, we first solve the regular (non-disordered) one-period W1 PCW. Typical parameters relevant to GaAs structures are considered: in-plane hexagonal pattern of circular holes whose radii are $70.9$~nm, lattice parameter $a=240$~nm, and slab thickness $d=150$~nm with dielectric constant $\epsilon=12.11$. The TE-like projected band
structure, associated with a regular supercell of dimensions $a\times5a\sqrt{3}$, is shown in Fig.~\ref{fig:bandsW1} for the first Brillouin zone of the periodic W1 PCW. The band-edge of the fundamental W1 mode, which is the focus of the present work, is located at $932$~nm. The photonic dispersion has been calculated employing the GME approach using 243 reciprocal lattice vectors $\mathbf{G}$, tested for convergence, within a circle of radius $a|\mathbf{G}|_{\rm max}=19$ centered at $\mathbf{G}=0$ in the reciprocal space of the regular supercell. Since we are interested in a frequency region below the second order guided mode, we consider only one guided mode in the GME expansion. We have verified that the second-order guided mode introduces a small correction of about $0.1\%$ in the band-edge frequency of the fundamental W1 guided mode, and consequently it can be safely neglected. By considering one guided mode into the GME method we are able to solve the present fully three-dimensional system with the computational effort of a two-dimensional calculation \cite{andreani}. In addition, the real part of the quasi-normal mode frequencies, associated to the bands above the light line of the system, are naturally obtained using the GME approach. 

Figure~\ref{fig:bandsW1} determines the Bloch mode basis to be considered in the BME calculation; specifically, we use the four bands highlighted with orange points in the figure, which correspond to the dominant terms in the expansion of Eq.~(\ref{bmeexpansion}) near to the band-edge of the fundamental W1 mode, when disorder is introduced in the system \cite{savona1}. In the present paper, we focus in a disordered waveguide of length $100a$ (disordered supercell dimension of $100a\times5a\sqrt{3}$), which automatically determines a complete set of 100 $k$ points in the Brillouin zone of Fig.~\ref{fig:bandsW1}. We choose this set to be formed by the centers of the disordered-W1 reciprocal lattice unit cells, which have length $2\pi/(100a)$, as it is represented by the orange points along the four Bloch basis bands in Fig.~\ref{fig:bandsW1}.\\ 

In order to study optical Anderson localization modes, we introduce disorder in the position $(x^{(0)}_m,y^{(0)}_m)$ and radius $r^{(0)}_m$ of the $m$-th hole, by considering random fluctuations $\delta$ with Gaussian probability
\begin{align}\label{fluctuations}
(x_m,y_m)&=(x^{(0)}_m+\delta x,y^{(0)}_m+\delta y),\nonumber  \\
r_m&=r^{(0)}_m+\delta r,
\end{align}
where $(x_m,y_m)$ and $r_m$ are the fluctuated position and radius of the $m$-th hole. The standard deviation of the random fluctuations in position, $\sigma_p=\sigma_x=\sigma_y$, and radius (size), $\sigma_s$, are taken as our disorder parameters. Since PCs are also subject to intrinsic disorder, i.e., unintentionally introduced in the fabrication process, the total amount of disorder that will be considered in the system is \cite{Garcia1}
\begin{equation}\label{sigmas}
\sigma^2=\sigma^2_i+\sigma^2_e,
\end{equation}
where $\sigma_i$ is the magnitude of intrinsic disorder and $\sigma_e$ is either $\sigma_p$ or $\sigma_s$. Intrinsic disorder can be introduced by fluctuating either the hole positions \cite{Garcia1,nishan1} or sizes \cite{dario1}, or even both \cite{savona1,minkov1}, and more sophisticated models can be also introduced to take into account the hole roughness \cite{nishan2,minkov2}. We consider fluctuations in the hole positions as the main intrinsic disorder contribution and  set $\sigma_i=0.005a$ in all our calculations, corresponding to the state-of-art in GasAs/InGaAs fabrication techniques \cite{Garcia1}. The modelling of intrinsic disorder is not so important here, as our main focus is on the role of the extrinsic disorder.\\

We solve the $400\times400$ Hermitian eigenvalue problem of Eq.~(\ref{maxwelldis}) for each statistical realization of the disordered system, and only the modes whose localization lengths  smaller than $25a$ and that are within the Lifshitz tail of the fundamental waveguide mode are considered in the statistical analysis, corresponding to the disorder-induced Anderson-localized states. The localization length of the mode $L_\beta$ is calculated by using the inverse participation number\cite{savona1}
\begin{equation}\label{invpn}
L_\beta=\left[\int_{\rm s. cell}|\mathbf{H}_{\beta}(x,y=0,z=0)|^4dx\right]^{-1},
\end{equation}
which has units of length if we compute first the normalization integral $\int_{\rm s. cell}|\mathbf{H}_{\beta}(x,y=0,z=0)|^2dx=N$, and use the re-scaled magnetic field $\mathbf{H}_{\beta}\rightarrow\mathbf{H}_{\beta}/\sqrt{N}$. Employing the cutoff condition $L_{\rm max}=25a$ we avoid finite-size effects coming from the Bloch periodic condition of the disordered supercell. In all the results that follow,  we  consider 20 statistical realizations, and have verified that such number of instances is enough to achieve the convergence of the first and second moments of the computed distributions.

\section{Statistics of disorder-induced quality factors and mode volumes}\label{SQV}

\begin{figure}[t]
  \begin{center}
    \includegraphics[width=0.48\textwidth]{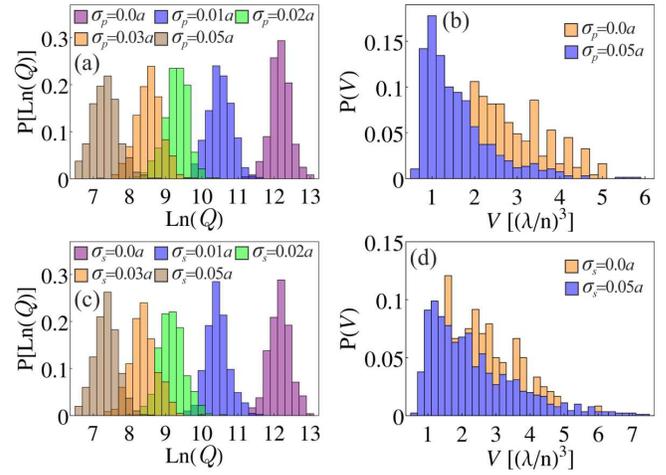}
  \end{center}
   \vspace{-0.3cm}
  \caption{
(Color online) Histograms of the natural logarithm of the quality factors, panel (a), and histograms of the mode volumes in units of $(\lambda/n)^3$, panel (b), for the Anderson-localized modes in the band-edge region of the fundamental waveguide mode, by considering disorder in the hole positions of the PC. Panels (c) and (d) correspond to the same histograms of (a) and (b), respectively, by considering disorder in the hole radii of the PC. In all cases, we also include an intrinsic disorder contribution of $\sigma_i=0.005a$.}
  \label{fig:QVH}
\end{figure}

Figure~\ref{fig:QVH} shows the quality factor and mode volume distributions of the Anderson-localized modes for different values of $\sigma_p$ [panels (a) and (b)] and $\sigma_s$ [panels (c) and (d)].
In all cases, there is also an intrinsic disorder contribution of $\sigma_i=0.005a$
as described above.
 The peak-position of the $Q$ factor distributions is displaced to smaller values as the degree of disorder increases, which was already addressed in theoretical \cite{lodahl1} and experimental works \cite{lodahl2}; moreover, the width of the distributions in the natural-logarithm histograms of Figs.~\ref{fig:QVH}~(a)~and~\ref{fig:QVH}(b) is approximately constant, which suggests that the standard deviation decreases when the magnitude of disorder increases. The peak and width of the $Q$ distributions display similar $\sigma_e$-dependence for the two models of disorder considered in this work. The behavior of the modal volume distributions, shown in Figs.~\ref{fig:QVH}~(b)~and~\ref{fig:QVH}(d), is however quantitatively different from the $Q$ distributions, since the volumes are concentrated in the small-volume region for large $\sigma_p$, panel (b), and slightly affected by large values of $\sigma_s$, panel (d). 

\begin{figure}[tp!]
  \begin{center}
    \includegraphics[width=0.48\textwidth]{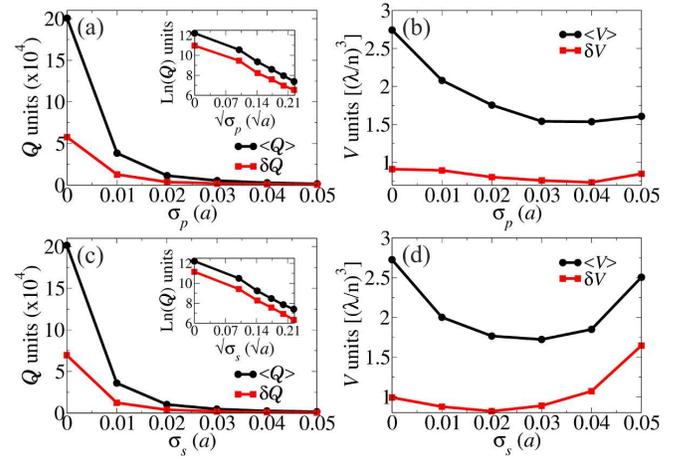}
  \end{center}
   \vspace{-0.3cm}
  \caption{(Color online) Mean, $\braket{Q}$,  and standard deviation, $\delta$Q, of the quality factors as a function of $\sigma_p$, panel (a), where the inset show the natural logarithm of $\braket{Q}$ and $\delta Q$ as a function of $\sqrt{\sigma_p}$. The corresponding mean, $\braket{V}$, and standard deviation, $\delta$V, of the mode volumes are shown in panel (b). Panels (c) and (d) correspond to the same analysis shown in panels (a) and (b), respectively, by considering disorder in the hole radii of the PC with disorder parameter $\sigma_s$. The lines only connect the individual points and serve as a guide for the eye.}
  \label{fig:QV}
\end{figure}

The explicit dependence of the mean value and standard deviation on the disorder parameter for $Q$ and $V$ distributions are shown in Fig.~\ref{fig:QV}. Panel \ref{fig:QV}(a) shows a fast decreasing of the mean quality factor, $\braket{Q}$, when $\sigma_p$ increases; such behaviour is also present in the standard deviation of the quality factor distribution, $\delta$Q, as  is suggested by the results in Fig.~\ref{fig:QVH}(a). In the inset of Fig.~\ref{fig:QV}(a), we shown the natural logarithm of  $\braket{Q}$ and $\delta Q$ as a function of $\sqrt{\sigma_p}$, where we clearly see a linear decreasing of these parameters as $\sqrt{\sigma_p}$ increases for deliberated disorder magnitudes larger than 1\% of the lattice parameter; thus we have an exponential dependence of both, $\braket{Q}$ and $\delta$Q, on the square root of the disorder parameter. The system displays the same qualitative behaviour by considering disorder in the hole radii, Fig.~\ref{fig:QV}(c), with approximately the same exponential constants found in the case of $\sqrt{\sigma_p}$ disorder for $\braket{Q}$ and $\delta$Q. We have also identified such a linear behaviour for $\sigma_p$ and $\sigma_s$ values smaller than 1\% (results not shown here); nevertheless, when the degree of extrinsic disorder approaches  zero, the localization length is comparable to the length of the disordered supercell considered in this work, leading to unphysical results due to finite-size effects. However, this numerical artifact can be avoided by considering a much larger disordered waveguide; for extrinsic disorder magnitudes of the order of 0.005$a$ or larger, relevant for deliberately disordered samples \cite{lodahl2,Garcia1,nishan1}, a waveguide length of $100a$ is sufficient to describe the main physics of the Anderson-localized modes. The mean of the effective mode volumes, $\braket{V}$, and their standard deviation, $\delta$V, as a function of the disorder magnitude are shown in Fig.~\ref{fig:QV}(b)~and~\ref{fig:QV}(d) for $\sigma_p$ and $\sigma_s$, respectively. 

\begin{figure}[tp!]
  \begin{center}
    \includegraphics[width=0.48\textwidth]{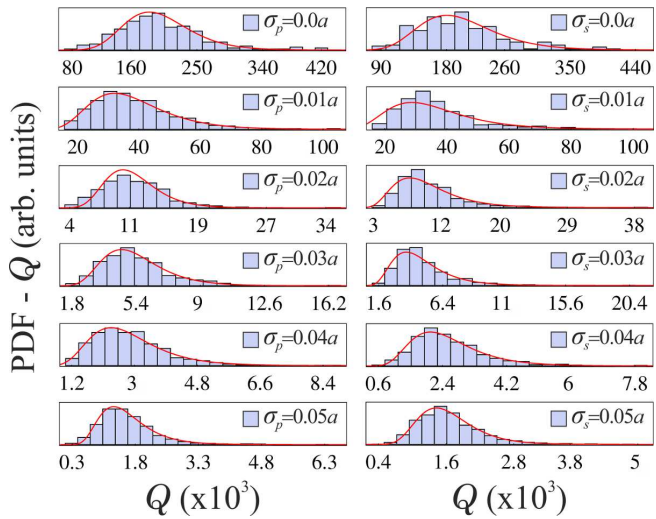}
  \end{center}
   \vspace{-0.3cm}
  \caption{(Color online) Probability density function (PDF) for the quality factor of the Anderson-localized modes in the region of the fundamental waveguide mode band-edge. Left panels correspond to deliberated $\sigma_p$ disorder and right panels correspond to deliberated $\sigma_s$ disorder.  In all cases, we also include an intrinsic disorder contribution of $\sigma_i=0.005a$. The continuous red curves are the corresponding log-normal distributions calculated with $\braket{Q}$ and $\delta$Q.}
  \label{fig:PDFQ}
\end{figure}

Our results show that the fluctuations in the mode volume are much smaller than the fluctuations in the quality factors, and a lower bound in the value of $\braket{V}$ as suggested already theoretically and experimentally  in recent works \cite{lalanne2}. Furthermore, in the region of large $\sigma_e$, and in contrast to the quality factor statistics, the behavior of $\braket{V}$ and $\delta$V is slightly different for both models of disorder; when deliberate disorder in the hole radii of the PC is considered the mean and the standard deviation of $V$ display an prominent minimum in the studied region of $\sigma_e$ [Fig.~\ref{fig:QV}(b)], while they display a much more soft behaviour when deliberated disorder in the hole positions is introduced in the system [Fig.~\ref{fig:QV}(d)]. These interesting results for the mode volume statistics suggest that in both models of disorder the localization length begins to increase when $\sigma_e$ is larger than 3\% of the lattice parameter, which is in agreement with the experimental work of Ref.~\onlinecite{lodahl2}. Nevertheless, between $\sigma_e=0$ and $\sigma_e=0.03a$, the mode volume decreases with increasing disorder, leading to a localization length decreasing as intuitively expected and reported in the literature \cite{mazoyer}. 

The symmetric Gaussian-like distributions in the log-histograms of Figs.~\ref{fig:QV}(a)~and~(c), demonstrate a log-normal distribution in the quality factor statistics. In order to clarify this trend, we plot the $Q$ probability density function (PDF) in Fig.~\ref{fig:PDFQ}, obtained from the Anderson-localized modes for all the statistical instances, superimposed with a log-normal distribution (red-solid curve) whose mean and standard deviation are calculated using $\braket{Q}$ and $\delta Q$. The left panels correspond to extrinsic $\sigma_p$ disorder, while the right panels correspond to extrinsic $\sigma_s$ disorder. When deliberate disorder in the hole positions is considered in the system, the quality factors statistics are very well described by a log-normal distribution, which has been assumed in previous works \cite{lodahl2} but, from our knowledge, never explicitly calculated with a fully three-dimensional model. Interestingly, when intermediate deliberated disorder magnitudes are introduced in the hole radii of the PC, the distributions of $Q$ are slightly different from the log-normal distribution, which suggests that, in the intermediate region of $\sigma_e$, the quality factors are differently distributed when different models of deliberated disorder are considered. These small deviations from the log-normal distributions in the $Q$ statistics will be focus of future works.

\section{Density of states and statistics of Purcell enhancement}\label{DOSPE}

One of the main advantages of the BME method, briefly discussed in Sec.~\ref{BME}, is the possibility to compute the out-of-plane losses of the disordered modes as well as their field profiles, allowing the calculation of the LDOS and PE using the Green's function formalism. To accomplish this, we adopt the following  mode expansion of the transverse Green's function \cite{yao}
\begin{equation}\label{Gfunphc}
\overleftrightarrow{\mathbf{G}}(\mathbf{r},\mathbf{r}',\omega)=\sum_\beta \frac{\omega^2\mathbf{E}_\beta^\ast(\mathbf{r}')\mathbf{E}_\beta(\mathbf{r})}{\tilde{\omega}_\beta^2-\omega^2},
\end{equation}
with the electric field subject to the normalization condition of Eq.~(\ref{enorm}). The Green's function of Eq.~(\ref{Gfunphc}) is solution of the point-source differential equation
\begin{equation}\label{Geq}
\left[\nabla\times\nabla\times-\frac{\omega^2}{c^2}
\epsilon({\bf r})
\right]\overleftrightarrow{\mathbf{G}}(\mathbf{r},\mathbf{r}',\omega)=
\frac{\omega^2}{c^2}\overleftrightarrow{\mathbf{I}}\delta(\mathbf{r}-\mathbf{r}'),
\end{equation}
where $\overleftrightarrow{\mathbf{I}}$ is the unit dyadic. The projected LDOS in the $\hat{n}_\mu$ direction is computed via \cite{novotny}
\begin{equation}\label{uLDOS}
\rho_\mu(\mathbf{r},\omega)=\frac{6}{\pi \omega}\left[\hat{n}_\mu\cdot\mbox{Im}\left\{\overleftrightarrow{\mathbf{G}}(\mathbf{r},\mathbf{r},\omega)\right\}\cdot\hat{n}_\mu\right],
\end{equation}
and the projected PE can be defined by
\begin{equation}\label{uPE}
\mbox{PE}_\mu(\mathbf{r},\omega)=\frac{\rho_\mu(\mathbf{r},\omega)}{\rho_\mu^0(\mathbf{r},\omega)}=\frac{\hat{n}_\mu\cdot\mbox{Im}\left\{\overleftrightarrow{\mathbf{G}}(\mathbf{r},\mathbf{r},\omega)\right\}\cdot\hat{n}_\mu}{\hat{n}_\mu\cdot\mbox{Im}\left\{\overleftrightarrow{\mathbf{G}}^0(\mathbf{r},\mathbf{r},\omega)\right\}\cdot\hat{n}_\mu},
\end{equation}
where \cite{patterson2}
\begin{equation}\label{prohom}
\hat{n}_\mu\cdot\mbox{Im}\left\{\overleftrightarrow{\mathbf{G}}^0(\mathbf{r},\mathbf{r},\omega)\right\}\cdot\hat{n}_\mu=\frac{\omega^3\sqrt{\epsilon}}{6\pi c^3},
\end{equation}
is the corresponding projection of the imaginary part of the Green's tensor in the bulk semiconductor, with dielectric constant $\epsilon$, solution of Eq.~(\ref{Geq}). From Eqs.~(\ref{Gfunphc}), (\ref{uPE}) and (\ref{prohom}), the PE factor at the position $\mathbf{r}$ and frequency $\omega$, projected in the $\hat{n}_\mu$ direction, reads
\begin{equation}\label{uPE2}
\mbox{PE}_\mu(\mathbf{r},\omega)=\frac{6\pi c^3}{\omega\sqrt{\epsilon}}\mbox{Im}\left\{\sum_\beta\frac{|E_{\mu,\beta}(\mathbf{r})|^2}{\tilde{\omega}_\beta^2-\omega^2}\right\}.
\end{equation}
The sum of the three orthogonal contributions of Eq.~(\ref{uPE2}) reduces to the well known result for the Purcell factor $[3/(4\pi^2)](\lambda/n)^3(\mbox{Q}_\beta/\mbox{V}_\beta)$, given $n=\sqrt{\epsilon}$, at $\omega=\omega_\beta$ in the electric field peak for low-loss photonic eigenmodes. The DOS can be calculated by tracing the Green's tensor over the three orthogonal spatial directions (associated to the total LDOS) and integrating its imaginary part over all space. In units of PE, i.e., over the density of states of the bulk semiconductor $\omega^2\sqrt{\epsilon}/(\pi^2 c^3)$, the DOS takes the following simple form
\begin{equation}\label{DOS}
\mbox{DOS}(\omega)=\frac{6\pi c^3}{\omega\epsilon^{3/2}}\mbox{Im}\left\{\sum_\beta\frac{1}{\tilde{\omega}_\beta^2-\omega^2}\right\},
\end{equation}
where we have used the normalization condition of Eq.~(\ref{enorm}) to compute the electric field integral.

\begin{figure}[t!]
  \begin{center}
    \includegraphics[width=0.48\textwidth]{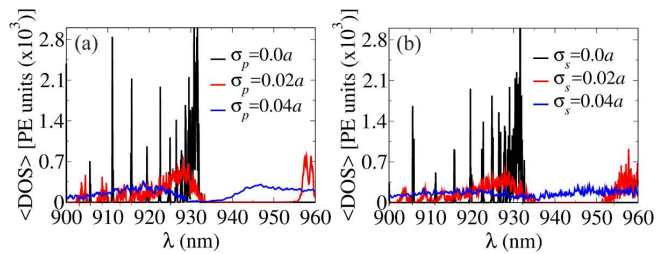}
  \end{center}
   \vspace{-0.3cm}
  \caption{(Color online) Averaged DOS over 20 statistical realizations, in PE units, by considering deliberated disorder in the hole positions of the PC, panel (a), and by considering deliberated disorder in their hole radii, panel (b). The intrinsic disorder contribution in all cases is  $\sigma_i=0.005a$.}
  \label{fig:DOS}
\end{figure}

Figure~\ref{fig:DOS} shows the computed averaged DOS, over 20 statistical realizations, by considering three values of $\sigma_p$, panel (a), and $\sigma_s$, panel (b), disorder magnitudes. Since the DOS is a property of the whole spectrum of the system we have used all the solutions coming from Eq.~(\ref{eigenBME}) in the region displayed in the figure, i.e., we have not considered any cutoff condition in the localization length of the eigenmodes. The characteristic Lifshitz tail is clearly evidenced and we capture the disorder-induced broadening of the DOS and the disorder-induced blue-shift of its peak, as already experimentally and theoretically reported (the latter using a perturbative approach with local field corrections) \cite{nishan1}. In the present analysis, the behavior of the PC band-edge (right peak) is also captured, due to the inclusion of this band in the Bloch mode basis (see the lowest frequency red band in Fig.~\ref{fig:bandsW1}). Notice that the Van Hove singularity at the band-edge of the fundamental waveguide mode is not present for $\sigma_p=0$ or $\sigma_s=0$ since we are considering a nominal amount of intrinsic disorder in the system. Our results show that when extrinsic disorder is considered in the hole radii of the PC, the DOS is slightly more affected than in the case when disorder in the hole position is deliberately introduced in the system. 

\begin{figure}[t!]
  \begin{center}
    \includegraphics[width=0.48\textwidth]{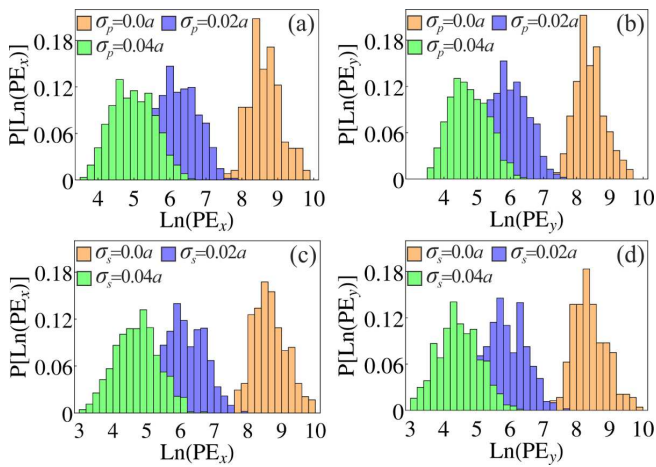}
  \end{center}
   \vspace{-0.3cm}
  \caption{(Color online) Histograms of the natural logarithm of the projected PEs in the $\hat{x}$, panel (a), and $\hat{y}$, panel (b), directions, associated to the corresponding electric field component peaks of the Anderson-localized modes, within the Lifshitz tail, by considering $\sigma_p$ disorder. Panels (c) and (d) correspond to the same histograms of (a) and (b), respectively, by considering $\sigma_s$ disorder. The intrinsic disorder contribution in all cases is  $\sigma_i=0.005a$.}
  \label{fig:PFH}
\end{figure}

\begin{figure*}[t!]
  \begin{center}
    \includegraphics[width=0.98\textwidth]{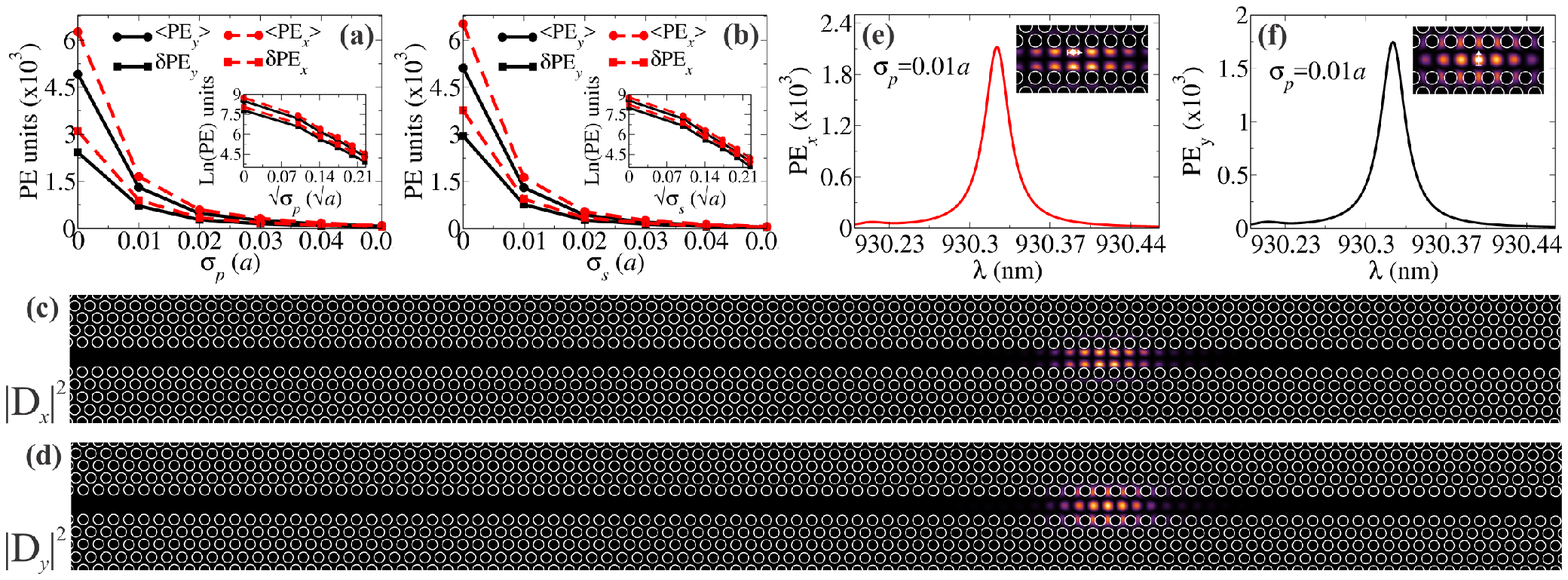}
  \end{center}
   \vspace{-0.3cm}
  \caption{(Color online) Mean, $<$PE$>$,  and standard deviation, $\delta$PE, of the $x$ and $y$ PEs as a function of $\sigma_p$, panel (a), and as a function of $\sigma_s$, panel (b). The corresponding insets show their natural logarithms as a function of the square of root of the extrinsic disorder parameter. The intensity profiles of the displacement field, $D_x$ and $D_y$ components, associated to an Anderson photonic mode, are shown in panels (c) and (d), respectively, for one statistical instance of the case $\sigma_p=0.01a$. Panels (e) and (f) show the PE at the peaks of $D_x$ and $D_y$, respectively, from panels (c) and (d), as a function of $\lambda$. The lines in panels (a) and (b) only connect the individual points and serve as a guide for the eye.}
  \label{fig:Purcell}
\end{figure*}

The natural-logarithm histograms of the $\hat{x}$ and $\hat{y}$ projections (the waveguide is along $x$ direction) of the PE, Eq.~(\ref{uPE2}), evaluated at the disorder mode frequencies and at the corresponding electric field component peaks, are shown in Fig.~\ref{fig:PFH} by considering $\sigma_p$ [panels (a) and (b)] and $\sigma_s$ [panels (c) and (d)] extrinsic disorder models. We find very broad PE distributions of the Anderson-localized modes within the Lifshitz tail, which are left-displaced for increasing degree of intentional disorder. The mean and standard deviation of the histograms are plotted in Figs.~\ref{fig:Purcell}(a)~and~\ref{fig:Purcell}(b) as a function of $\sigma_p$ and $\sigma_s$, respectively. Both $\hat{x}$ and $\hat{y}$, projections of the mean PE display a decreasing behaviour when the extrinsic amount of disorder increases. The same trend is seen in the standard deviation of the distributions leading to a concentration of the enhancement for large deliberated disorder magnitudes. From Figs.~\ref{fig:Purcell}(a)~and~\ref{fig:Purcell}(b), it is also clear that the enhancement factor is larger in the $\hat{x}$ direction (along the waveguide) than in $\hat{y}$ direction (perpendicular to the waveguide in the PC plane), which is expected given the slightly larger peak of the $E_x$ component in comparison to the peak of  $E_y$. Our results demonstrate that the positions of the $E_x$ peaks are therefore the best places for cavity-QED applications, as it will be discussed in Sec.~\ref{SC}. The system also displays approximately the same behaviour for both models of extrinsic disorder. In the insets of Figs.~\ref{fig:Purcell}(a)~and~\ref{fig:Purcell}(b) we show the natural logarithm of $<$PE$>$ and $\delta$PE for hole position and hole size disorder, respectively, as a function of the square root of the intentional disorder parameter. We find the same linear behavior seen in the corresponding insets for the quality factor analysis presented in Figs.~\ref{fig:QV}(a)~and~\ref{fig:QV}(b): an exponential dependence of the PE projections on the square root of the extrinsic disorder parameter, $\sqrt{\sigma_e}$, with approximately the same exponential constants found for the quality factors in the case of $\sigma_p$, and slightly smaller than the corresponding exponential constants found for them in the case of $\sigma_s$. The small difference between the exponential constants for these two different disorder models appears due to the slightly different response of DOS to hole position and size disorder (see Fig.~\ref{fig:DOS}). However, since the mean of the mode volumes changes within a much less broad region, in $(\lambda/n)^3$ units, in comparison to the quality factors, where the means change about two orders of magnitude for the $\sigma_e$ values considered (see Fig.~\ref{fig:QV}), the PEs, which are roughly proportional to $Q/V$, are mainly determined by the $Q$ statistics of the Anderson-localized modes, which explains why the PE always decreases even for decreasing $V$. Future works could focus on the regime where the $Q$ and $V$ variation ranges are comparable, i.e., very large intentional disorder, in which the dependence of PE on $\sigma_e$ is determined by both, $Q$ and $V$ statistics. 

\begin{figure}[t!]
  \begin{center}
    \includegraphics[width=0.48\textwidth]{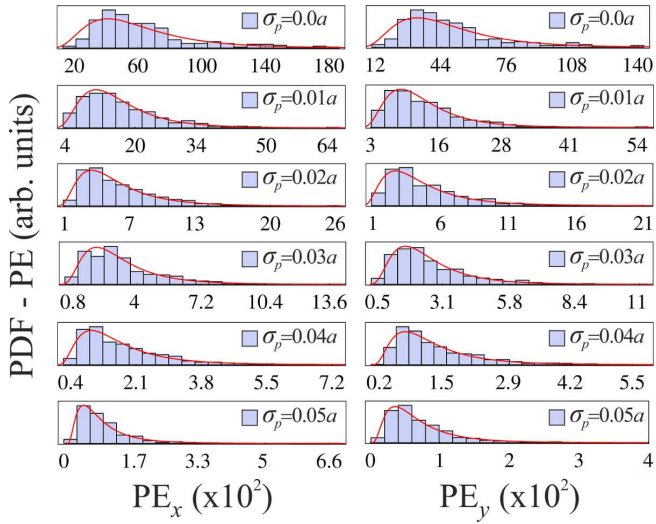}
  \end{center}
   \vspace{-0.3cm}
  \caption{(Color online) Probability density function of the PEs in the $\hat{x}$, left panels, and $\hat{y}$, right panels, directions when deliberated disorder is introduced in the hole positions of the PC.  The continuous red curves are the corresponding log-normal distributions calculated with $<$PE$>$ and $\delta$PE. The intrinsic disorder contribution in all cases is  $\sigma_i=0.005a$.}
  \label{fig:PFpPDF}
\end{figure}

\begin{figure}[t!]
 \begin{center}
   \includegraphics[width=0.48\textwidth]{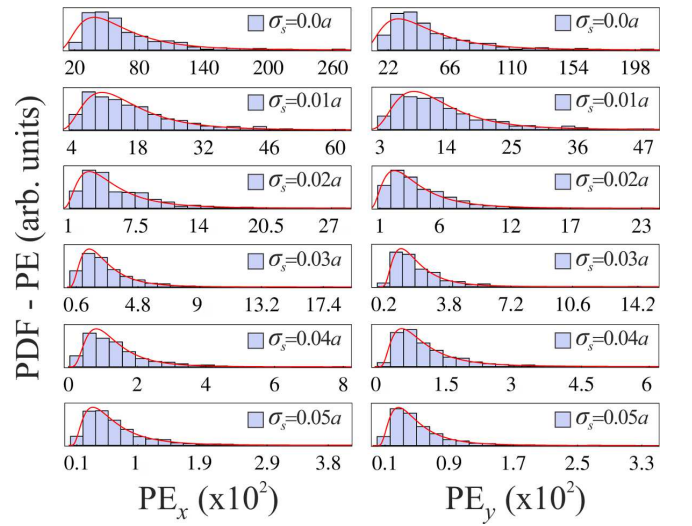}
 \end{center}
  \vspace{-0.3cm}
  \caption{(Color online) Probability density function of the PEs in the $\hat{x}$, left panels, and $\hat{y}$, right panels, directions when deliberated disorder is introduced in the hole radii of the PC. The continuous red curves are the corresponding log-normal distributions calculated with $<$PE$>$ and $\delta$PE.  The intrinsic disorder contribution in all cases is  $\sigma_i=0.005a$.}
  \label{fig:PFsPDF}
\end{figure}

It is important to realise that the variation range of the Anderson-localized mode volumes is strongly dependent on the waveguide length and consequently on the cutoff condition $L_{\rm max}$. Since the change in the mode volume is mainly determined by the change in the localization length and we are within a narrow region of the spectrum, the former is roughly proportional to the latter $V_\beta\propto L_\beta$; then, given a $V$ variation of one $(\lambda/n)^3$ unit and taking into account from our calculations that the localization length varies from $\sim6a$ to $L_{\rm max}=25a$, we estimate a variation of $\sim20(\lambda/n)^3$ units for $V$ by considering a $L_{\rm max}=416a$ cutoff, corresponding to a typical waveguide length, i.e., $100~\mu$m. Therefore, we believe that the $Q$ statistics also dominate the behavior of the PE for sate-of-art samples, well into the disorder regime investigated in the present work. The intensity profiles of the displacement field [$\mathbf{D}(\mathbf{r})=\epsilon'(\mathbf{r})\mathbf{E}(\mathbf{r})$] components  $D_x$ and $D_y$, for an Anderson-localized mode with $\sigma_p=0.01a$, are shown in Figs.~\ref{fig:Purcell}(c)~and~\ref{fig:Purcell}(d), respectively, where the strong localization in the waveguide region is clearly seen. We compute the PE for this mode at the corresponding electric field component peaks in function of lambda, and we show the results in panels (e), $\hat{x}$ projection, and (f), $\hat{y}$ projection, of Fig.~\ref{fig:Purcell}. As it is expected, the enhancement is magnified at resonance with the Anderson localized photonic mode. Figures~\ref{fig:PFpPDF}~and~\ref{fig:PFsPDF} show the PDF of PE when deliberated disorder is introduced in the hole positions and hole radii, respectively, and the red curve is the corresponding log-normal distribution determined by $<$PE$>$ and $\delta$PE. Given the Gaussian-like histograms of Fig.~\ref{fig:PFH}, whose behaviour is dominated by the quality factors of the Anderson-localized modes, the PE statistics is therefore well described by log-normal distributions in the considered extrinsic disorder regime.

The present calculations on the statistics of PEs, as already discussed, were carried out by considering optimal spectral and spatial conditions, i.e., Eq.~(\ref{uPE2}) is evaluated at the frequency of the Anderson-localized mode (zero detuning) and at the electric field component peak. Consequently, our results determine upper bounds on the PEs that can be measure in state-of-art W1 PCW samples, for which lower bounds have already been experimentally established \cite{lodahl3}.

\section{Probability of strong coupling with a quantum emitter} \label{SC}

The possibility to achieve the strong coupling regime between the Anderson-localized photonic modes and a single quantum emitter has been partly theoretically addressed by Henri \textit{et al.} \cite{lodahl1} using simple one-dimensional disordered PCs, with experimental signatures  experimentally observed by  Gao \textit{et al.} \cite{wong} in the slow-light regime. Furthermore,  Caz\'e \textit{et al.} \cite{carminati} studied the strong coupling criteria with two-dimensional Anderson-localized modes. Here, we compute numerically the probability of strong coupling using the criterion obtained by  Caz\'e \textit{et al.}, and consider, in a natural and realistic way, the cavity loss spatial distributions determined by the quality factor statistics; thus we generalize the simpler model result obtained by Henri \textit{et al.} where the same loss length is assumed for all Anderson-localized modes. Indeed, Smolka \textit{et al.} \cite{lodahl2} discussed the importance of including a loss length distribution in order to obtain accurate statistical results. Using a quantum mechanical theory of radiation-matter interaction, the coupling constant between a quantum emitter and an electromagnetic eigenmode, in the dipole approximation, is given as \cite{hughes1} 
\begin{equation}\label{gdef}
g_\beta=\sqrt{\frac{\omega_\beta}{2\hbar\epsilon_0}}\mathbf{d}\cdot\mathbf{E}_\beta(\mathbf{r}_{d}),
\end{equation}

\noindent where $\mathbf{r}_{d}$ and $\mathbf{d}=d\hat{e}$ are the position and the $\hat{e}$-oriented dipole moment, respectively, of the quantum dipole emitter. Here, we consider dipole moments of $d=40$~Debye, relevant to self-organized InGaAs QDs \cite{stievater,jrguest,minkov4}. By assuming the QD mimics a simple two-level system in the frequency regime of interest, we adopt the approximate condition for strong coupling \cite{carminati}   
\begin{equation}\label{SCcond}
g_\beta^2\geq\frac{\gamma_\beta^2+\gamma_{d}^2}{8},
\end{equation}
where $\gamma_\beta=\omega_\beta/Q_\beta$ is the photonic cavity decay rate and $\gamma_{d}$ represents an overall exciton decay rate in the quantum dot exciton (without cavity coupling), including the rate of radiative decay into the photonic modes and any possible non-radiative mechanism. Equation~(\ref{SCcond}) ensures that the Rabi splitting is larger than the polariton linewidth, which is a necessary and sufficient condition to observe the frequency splitting between the polariton states, and it is more restrictive than the usual adopted condition $g_\beta^2\geq(\gamma_\beta^2 -\gamma^2_d)/16$, which ensures only the existence of a non-zero Rabi splitting between the exciton-photon coupled modes \cite{reithmaier,andreanisc}. We then define the quantity 
\begin{equation}\label{SCf}
\mbox{SC}_\beta=g_\beta^2-\frac{\gamma_\beta^2+\gamma_{d}^2}{8},
\end{equation}
and calculate the probability that SC$_\beta$ is larger than zero within the sample space defined by all the statistical realizations of the disordered system, which is merely the probability that the quantum emitter is strongly coupled to an Anderson-localized photonic mode. We specifically consider $\mathbf{r}_{d}=(x_{d},y_{d},0)$, which is the region where the field is more intense for our TE-like polarization, and separate the contributions of $E_x$ and $E_y$ electric field components in Eq.~(\ref{gdef}) (the $E_z$ is zero at $z=0$), which is equivalent to consider dipole orientations in either $\hat{x}$ or $\hat{y}$ directions, respectively. Since the PC is randomly perturbed when disorder is introduced, the probability of field localization must be the same along the waveguide for Anderson modes, and hence the probability of strong coupling should be independent on the emitter position along this region. Nevertheless, special care must be taken into account since the localization positions of the field antinodes do not vary continuously along the waveguide and the strength of the dot-field interaction depends on the dipole orientation of the quantum emitter. 

\begin{figure}[t!]
  \begin{center}
    \includegraphics[width=0.48\textwidth]{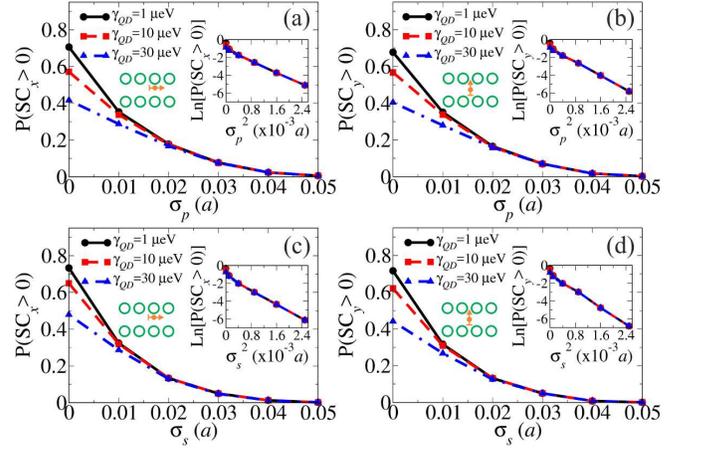}
  \end{center}
   \vspace{-0.3cm}
  \caption{(Color online) Probability of a $d=40~$Debye QD exciton with overall exciton decay rate of $\gamma_{d}=1~\mu$eV (black-circles), $\gamma_{d}=10~\mu$eV (red-squares) and $\gamma_{d}=30~\mu$eV (blue-triangles) be strongly coupled (SC) to an Anderson-localized mode. Panels (a) and (b) corresponds to dipole moment orientations in the $\hat{x}$ and $\hat{y}$ directions, respectively, by considering $\sigma_p$ disorder. The intrinsic disorder contribution in all cases is  $\sigma_i=0.005a$. Panels (c) and (d) correspond to the same plots in panels (a) and (b), respectively, by considering $\sigma_s$ disorder. The insets show the natural logarithm of the strong coupling probability as a function of the squared extrinsic disorder parameter. The lines only connect the individual points and serve as a guide for the eye.}
  \label{fig:SC}
\end{figure}

From the insets of Figs.~\ref{fig:Purcell}(e)~and~\ref{fig:Purcell}(f) we clearly see that the antinodes of the $E_x$ component are positioned between the missing holes defining the waveguide and they are outward-displaced from the line $y=0$, while the antinodes of the $E_y$ component are positioned at approximately the same positions of such missing holes. We have verified this for all the Anderson-localized modes studied in the present work. Based on our numerical results, we then obtain the optimal regions and orientations for positioning the QD: between missing holes at $y=0.433a$ for a $\hat{x}$ oriented dipole, and at the missing holes with $y=0$ for a $\hat{y}$ oriented one. We choose one point of these optimal statistically equivalent regions for each orientation to compute the probability of strong coupling. In our particular coordinate system, where there is no missing hole at $(x=0,y=0)$, we have $(x_{d},y_{d})=(0,0.433a)$ and $(x_{d},y_{d})=(0.5a,0.0)$ for the $\hat{x}$ and $\hat{y}$ oriented dipoles, respectively. Results are shown in Fig.~\ref{fig:SC} for $\sigma_p$, panels (a) and (b), and $\sigma_s$, panels (c) and (d), extrinsic disorder parameters, where we have considered $\gamma_{d}=1~\mu$eV (black-circles), $\gamma_{d}=10~\mu$eV (red-squares) and $\gamma_{d}=30~\mu$eV (blue-triangles). The probability of strong coupling always decreases for increasing magnitudes of deliberated disorder, and it is slightly larger when the dipole moment is oriented in $\hat{x}$, i.e., along the waveguide direction (see inset scheme), which is a direct consequence of the $\hat{x}$ projected PE be larger than the $\hat{y}$ projected one (see Fig.~\ref{fig:Purcell}). We also note that the differences on the strong coupling probabilities for different values of $\gamma_{d}$ at the same $\sigma_e$, decreases when $\sigma_e$ increases, which is due to the dominant effect of $\gamma_\beta$ (low $Q_\beta$) for large disorder. This result suggest that state-of-art QDs, corresponding to the region between the red and black curves, may lead to large strong coupling probabilities only in the low disorder regime, around of 60\% for non-intentional disordered samples and around 30\% for an intentional disorder of the order of $0.01a$. Also note that these results correspond to upper bounds on the actual strong coupling probability since we are considering quantum dots with a mean overall rate $\gamma_{d}$, perfectly oriented, at resonance with the Anderson photonic mode and positioned at the field antinodes. The latter is not however a great limitation given the state-of-art precision on quantum dot positioning (around $20$~nm) in modern nano-fabrication techniques \cite{kevin07nat,reinhard}; and the resonant condition can be achieved by using the quantum confined Stark effect \cite{villasboas,faraon} in temperature-dependent measurements \cite{wong}. In the insets of Fig.~\ref{fig:SC}, we plot the natural logarithms of P(SC) as a function of $\sigma_e^2$, and we see that the probability of strong coupling is well described by a decreasing exponential behaviour on the square of the extrinsic disorder parameter. When disorder in the hole radii is considered, Figs.~\ref{fig:SC}(c)~and~\ref{fig:SC}(d), the exponential constants are found to be around 25\% larger than the case when the hole positions are deliberately disordered, Figs.~\ref{fig:SC}(a)~and~\ref{fig:SC}(b), therefore, the strong coupling probability is more robust when the latter model of intentional disorder is considered in the system. We attribute the difference in the exponential constants to the fact that the total DOS is more affected by $\sigma_s$ than by $\sigma_p$, as it can be seen from Fig.~\ref{fig:DOS}. 

\section{Conclusions} \label{Concl}

Using the fully three-dimensional Bloch mode expansion method we have numerically calculated the quality factor and mode volumes distributions of the Anderson-localized modes in deliberately disordered W1 photonic crystal slab waveguides. We have considered intentional disorder in the PCW by randomly fluctuating either its hole positions (standard deviation $\sigma_p$) or its hole radii (standard deviation $\sigma_s$) with Gaussian probability, and state-of-art intrinsic disorder by introducing additional Gaussian fluctuations in its hole positions with standard deviation $\sigma_i$. We find that the mean effective mode volume of the Anderson modes decreases with increasing small disorder degrees and increases with increasing large disorder degrees, displaying a lower bound as it was recently predicted for the localization length using two-dimensional calculations. Such a behavior is seen to be more accentuated when disorder in the hole sizes is considered. By studying the functional dependence of the mean and standard deviation of the quality factor distributions on $\sigma_p$ and $\sigma_s$, we have found that these statistical parameters decrease exponentially on the square root of the intentional disorder parameter. Furthermore, we have shown that the quality factor distributions are in general well described by log-normal distributions, without any fitting parameter, as already suggested in recent experimental works \cite{lodahl2}. Adopting a Green function approach, we also computed, for the first time to our knowledge, the statistical distributions of Purcell enhancements using the projected LDOS in the orthogonal $\hat{x}$ and $\hat{y}$ directions. The mean values of the Purcell enhancements found in this work, as a function of the extrinsic disorder parameter, represent upper bounds to the corresponding values that could be experimentally measured in the present system, for which lower bounds have already been addressed in recent experiments \cite{lodahl3}. Moreover, in the studied disorder regime, the statistics of the Purcell enhancement is mainly determined by the quality factor distributions, which is verified by the same exponential dependence of the enhancement on the square root of the intentional disorder parameter. As a consequence of the $Q$ dominant behaviour on the Purcell enhancement, we have found that the Purcell enhancement statistics is also well described by log-normal distributions. We have also discussed the possibility of strong coupling between a quantum emitter, positioned within the waveguide, and an Anderson-localized mode. We determined the best regions for positioning the quantum emitter and we found that the probability of strong coupling always decreases for increasing intentional disorder, where low disordered systems give rise to larger probabilities when state-of-art quantum dots are considered. We believe that these result will be of remarkably importance for guiding and motivating new experiments involving Anderson-localized modes with promising potentialities on cavity-QED \cite{sapienza}. Finally, we studied the functional dependence of the strong coupling probability on the intentional disorder parameter and we found an exponential decreasing on $\sigma_s^2$ and $\sigma_p^2$, where the exponential constants of the former disorder are larger than exponential constants of the latter one, i.e., the strong coupling probability is more robust against hole-position disorder than against hole-size disorder.

\section{Acknowledgements}

This work was supported by the Natural Sciences and Engineering Research Council of Canada (NSERC) and Queen's University, Canada. J.P. Vasco would like to thank Momchil Minkov for useful discussions on the BME method. This research was enabled in part by computational support provided by the Centre for Advanced Computing (http://cac.queensu.ca) and  Compute Canada (www.computecanada.ca).


\begin{thebibliography}{61}%
\makeatletter
\providecommand \@ifxundefined [1]{%
 \@ifx{#1\undefined}
}%
\providecommand \@ifnum [1]{%
 \ifnum #1\expandafter \@firstoftwo
 \else \expandafter \@secondoftwo
 \fi
}%
\providecommand \@ifx [1]{%
 \ifx #1\expandafter \@firstoftwo
 \else \expandafter \@secondoftwo
 \fi
}%
\providecommand \natexlab [1]{#1}%
\providecommand \enquote  [1]{``#1''}%
\providecommand \bibnamefont  [1]{#1}%
\providecommand \bibfnamefont [1]{#1}%
\providecommand \citenamefont [1]{#1}%
\providecommand \href@noop [0]{\@secondoftwo}%
\providecommand \href [0]{\begingroup \@sanitize@url \@href}%
\providecommand \@href[1]{\@@startlink{#1}\@@href}%
\providecommand \@@href[1]{\endgroup#1\@@endlink}%
\providecommand \@sanitize@url [0]{\catcode `\\12\catcode `\$12\catcode
  `\&12\catcode `\#12\catcode `\^12\catcode `\_12\catcode `\%12\relax}%
\providecommand \@@startlink[1]{}%
\providecommand \@@endlink[0]{}%
\providecommand \url  [0]{\begingroup\@sanitize@url \@url }%
\providecommand \@url [1]{\endgroup\@href {#1}{\urlprefix }}%
\providecommand \urlprefix  [0]{URL }%
\providecommand \Eprint [0]{\href }%
\providecommand \doibase [0]{http://dx.doi.org/}%
\providecommand \selectlanguage [0]{\@gobble}%
\providecommand \bibinfo  [0]{\@secondoftwo}%
\providecommand \bibfield  [0]{\@secondoftwo}%
\providecommand \translation [1]{[#1]}%
\providecommand \BibitemOpen [0]{}%
\providecommand \bibitemStop [0]{}%
\providecommand \bibitemNoStop [0]{.\EOS\space}%
\providecommand \EOS [0]{\spacefactor3000\relax}%
\providecommand \BibitemShut  [1]{\csname bibitem#1\endcsname}%
\let\auto@bib@innerbib\@empty
\bibitem [{\citenamefont {Yablonovitch}(1987)}]{yablonovitch}%
  \BibitemOpen
  \bibfield  {author} {\bibinfo {author} {\bibfnamefont {E.}~\bibnamefont
  {Yablonovitch}},\ }\href@noop {} {\bibfield  {journal} {\bibinfo  {journal}
  {Phys. Rev. Lett.}\ }\textbf {\bibinfo {volume} {58}},\ \bibinfo {pages}
  {2059} (\bibinfo {year} {1987})}\BibitemShut {NoStop}%
\bibitem [{\citenamefont {John}(1987)}]{john}%
  \BibitemOpen
  \bibfield  {author} {\bibinfo {author} {\bibfnamefont {S.}~\bibnamefont
  {John}},\ }\href@noop {} {\bibfield  {journal} {\bibinfo  {journal} {Phys.
  Rev. Lett.}\ }\textbf {\bibinfo {volume} {58}},\ \bibinfo {pages} {2486}
  (\bibinfo {year} {1987})}\BibitemShut {NoStop}%
\bibitem [{\citenamefont {Joannopoulos}\ \emph {et~al.}(2008)\citenamefont
  {Joannopoulos}, \citenamefont {Johnson}, \citenamefont {Winn},\ and\
  \citenamefont {Meade}}]{joannopoulos}%
  \BibitemOpen
  \bibfield  {author} {\bibinfo {author} {\bibfnamefont {J.~D.}\ \bibnamefont
  {Joannopoulos}}, \bibinfo {author} {\bibfnamefont {S.~G.}\ \bibnamefont
  {Johnson}}, \bibinfo {author} {\bibfnamefont {J.~N.}\ \bibnamefont {Winn}}, \
  and\ \bibinfo {author} {\bibfnamefont {R.~D.}\ \bibnamefont {Meade}},\
  }\href@noop {} {\emph {\bibinfo {title} {{Photonic Crystals: Molding the Flow
  of Light}}}},\ \bibinfo {edition} {2nd}\ ed.\ (\bibinfo  {publisher}
  {Princeton University Press},\ \bibinfo {year} {2008})\BibitemShut {NoStop}%
\bibitem [{\citenamefont {Reinhard}\ \emph {et~al.}(2012)\citenamefont
  {Reinhard}, \citenamefont {Volz}, \citenamefont {Winger}, \citenamefont
  {Badolato}, \citenamefont {Hennessy}, \citenamefont {Hu},\ and\ \citenamefont
  {Imamo\u{g}lu}}]{reinhard}%
  \BibitemOpen
  \bibfield  {author} {\bibinfo {author} {\bibfnamefont {A.}~\bibnamefont
  {Reinhard}}, \bibinfo {author} {\bibfnamefont {T.}~\bibnamefont {Volz}},
  \bibinfo {author} {\bibfnamefont {M.}~\bibnamefont {Winger}}, \bibinfo
  {author} {\bibfnamefont {A.}~\bibnamefont {Badolato}}, \bibinfo {author}
  {\bibfnamefont {K.~J.}\ \bibnamefont {Hennessy}}, \bibinfo {author}
  {\bibfnamefont {E.~L.}\ \bibnamefont {Hu}}, \ and\ \bibinfo {author}
  {\bibfnamefont {A.}~\bibnamefont {Imamo\u{g}lu}},\ }\href@noop {} {\bibfield
  {journal} {\bibinfo  {journal} {Nat. Photon.}\ }\textbf {\bibinfo {volume}
  {6}},\ \bibinfo {pages} {93} (\bibinfo {year} {2012})}\BibitemShut {NoStop}%
\bibitem [{\citenamefont {Yoshie}\ \emph {et~al.}(2004)\citenamefont {Yoshie},
  \citenamefont {Scherer}, \citenamefont {Hendrickson}, \citenamefont
  {Khitrova}, \citenamefont {Gibbs}, \citenamefont {Rupper}, \citenamefont
  {Ell}, \citenamefont {Shchekin},\ and\ \citenamefont {Deppe}}]{yoshie}%
  \BibitemOpen
  \bibfield  {author} {\bibinfo {author} {\bibfnamefont {T.}~\bibnamefont
  {Yoshie}}, \bibinfo {author} {\bibfnamefont {A.}~\bibnamefont {Scherer}},
  \bibinfo {author} {\bibfnamefont {J.}~\bibnamefont {Hendrickson}}, \bibinfo
  {author} {\bibfnamefont {G.}~\bibnamefont {Khitrova}}, \bibinfo {author}
  {\bibfnamefont {H.~M.}\ \bibnamefont {Gibbs}}, \bibinfo {author}
  {\bibfnamefont {G.}~\bibnamefont {Rupper}}, \bibinfo {author} {\bibfnamefont
  {C.}~\bibnamefont {Ell}}, \bibinfo {author} {\bibfnamefont {O.~B.}\
  \bibnamefont {Shchekin}}, \ and\ \bibinfo {author} {\bibfnamefont {D.~G.}\
  \bibnamefont {Deppe}},\ }\href@noop {} {\bibfield  {journal} {\bibinfo
  {journal} {Nature (London)}\ }\textbf {\bibinfo {volume} {432}},\ \bibinfo
  {pages} {200} (\bibinfo {year} {2004})}\BibitemShut {NoStop}%
\bibitem [{\citenamefont {Hennessy}\ \emph
  {et~al.}(2007{\natexlab{a}})\citenamefont {Hennessy}, \citenamefont
  {Badolato}, \citenamefont {Winger}, \citenamefont {Gerace}, \citenamefont
  {Atat{\"u}re}, \citenamefont {Gulde}, \citenamefont {F{\"a}lt}, \citenamefont
  {Hu},\ and\ \citenamefont {Imamo\v{g}lu}}]{hennessy}%
  \BibitemOpen
  \bibfield  {author} {\bibinfo {author} {\bibfnamefont {K.}~\bibnamefont
  {Hennessy}}, \bibinfo {author} {\bibfnamefont {A.}~\bibnamefont {Badolato}},
  \bibinfo {author} {\bibfnamefont {M.}~\bibnamefont {Winger}}, \bibinfo
  {author} {\bibfnamefont {D.}~\bibnamefont {Gerace}}, \bibinfo {author}
  {\bibfnamefont {M.}~\bibnamefont {Atat{\"u}re}}, \bibinfo {author}
  {\bibfnamefont {S.}~\bibnamefont {Gulde}}, \bibinfo {author} {\bibfnamefont
  {S.}~\bibnamefont {F{\"a}lt}}, \bibinfo {author} {\bibfnamefont {E.~L.}\
  \bibnamefont {Hu}}, \ and\ \bibinfo {author} {\bibnamefont {Imamo\v{g}lu}},\
  }\href@noop {} {\bibfield  {journal} {\bibinfo  {journal} {Nature (London)}\
  }\textbf {\bibinfo {volume} {445}},\ \bibinfo {pages} {896} (\bibinfo {year}
  {2007}{\natexlab{a}})}\BibitemShut {NoStop}%
\bibitem [{\citenamefont {Laucht}\ \emph {et~al.}(2010)\citenamefont {Laucht},
  \citenamefont {Villas-B{\^o}as}, \citenamefont {Stobbe}, \citenamefont
  {Hauke}, \citenamefont {Hofbauer}, \citenamefont {B{\"o}hm}, \citenamefont
  {Lodahl}, \citenamefont {Amann}, \citenamefont {Kaniber},\ and\ \citenamefont
  {Finley}}]{villasboas}%
  \BibitemOpen
  \bibfield  {author} {\bibinfo {author} {\bibfnamefont {A.}~\bibnamefont
  {Laucht}}, \bibinfo {author} {\bibfnamefont {J.~M.}\ \bibnamefont
  {Villas-B{\^o}as}}, \bibinfo {author} {\bibfnamefont {S.}~\bibnamefont
  {Stobbe}}, \bibinfo {author} {\bibfnamefont {N.}~\bibnamefont {Hauke}},
  \bibinfo {author} {\bibfnamefont {F.}~\bibnamefont {Hofbauer}}, \bibinfo
  {author} {\bibfnamefont {G.}~\bibnamefont {B{\"o}hm}}, \bibinfo {author}
  {\bibfnamefont {P.}~\bibnamefont {Lodahl}}, \bibinfo {author} {\bibfnamefont
  {M.-C.}\ \bibnamefont {Amann}}, \bibinfo {author} {\bibfnamefont
  {M.}~\bibnamefont {Kaniber}}, \ and\ \bibinfo {author} {\bibfnamefont
  {J.~J.}\ \bibnamefont {Finley}},\ }\href@noop {} {\bibfield  {journal}
  {\bibinfo  {journal} {Phys. Rev. B}\ }\textbf {\bibinfo {volume} {82}},\
  \bibinfo {pages} {075305} (\bibinfo {year} {2010})}\BibitemShut {NoStop}%
\bibitem [{\citenamefont {Yao}\ and\ \citenamefont {Hughes}(2009)}]{hughes2}%
  \BibitemOpen
  \bibfield  {author} {\bibinfo {author} {\bibfnamefont {P.}~\bibnamefont
  {Yao}}\ and\ \bibinfo {author} {\bibfnamefont {S.}~\bibnamefont {Hughes}},\
  }\href@noop {} {\bibfield  {journal} {\bibinfo  {journal} {Opt. Express}\
  }\textbf {\bibinfo {volume} {17}},\ \bibinfo {pages} {11505} (\bibinfo {year}
  {2009})}\BibitemShut {NoStop}%
\bibitem [{\citenamefont {Vasco}\ \emph {et~al.}(2016)\citenamefont {Vasco},
  \citenamefont {Gerace}, \citenamefont {Guimar{\~a}es},\ and\ \citenamefont
  {Santos}}]{jpvasco}%
  \BibitemOpen
  \bibfield  {author} {\bibinfo {author} {\bibfnamefont {J.~P.}\ \bibnamefont
  {Vasco}}, \bibinfo {author} {\bibfnamefont {D.}~\bibnamefont {Gerace}},
  \bibinfo {author} {\bibfnamefont {P.~S.~S.}\ \bibnamefont {Guimar{\~a}es}}, \
  and\ \bibinfo {author} {\bibfnamefont {M.~F.}\ \bibnamefont {Santos}},\
  }\href@noop {} {\bibfield  {journal} {\bibinfo  {journal} {Phys. Rev. B}\
  }\textbf {\bibinfo {volume} {94}},\ \bibinfo {pages} {165302} (\bibinfo
  {year} {2016})}\BibitemShut {NoStop}%
\bibitem [{\citenamefont {Galli}\ \emph {et~al.}(2010)\citenamefont {Galli},
  \citenamefont {Gerace}, \citenamefont {Welna}, \citenamefont {Krauss},
  \citenamefont {O{\rq}Faolain}, \citenamefont {Guizzetti},\ and\ \citenamefont
  {Andreani}}]{galli}%
  \BibitemOpen
  \bibfield  {author} {\bibinfo {author} {\bibfnamefont {M.}~\bibnamefont
  {Galli}}, \bibinfo {author} {\bibfnamefont {D.}~\bibnamefont {Gerace}},
  \bibinfo {author} {\bibfnamefont {K.}~\bibnamefont {Welna}}, \bibinfo
  {author} {\bibfnamefont {T.~F.}\ \bibnamefont {Krauss}}, \bibinfo {author}
  {\bibfnamefont {L.}~\bibnamefont {O{\rq}Faolain}}, \bibinfo {author}
  {\bibfnamefont {G.}~\bibnamefont {Guizzetti}}, \ and\ \bibinfo {author}
  {\bibfnamefont {L.~C.}\ \bibnamefont {Andreani}},\ }\href@noop {} {\bibfield
  {journal} {\bibinfo  {journal} {Opt. Express}\ }\textbf {\bibinfo {volume}
  {18}},\ \bibinfo {pages} {26613} (\bibinfo {year} {2010})}\BibitemShut
  {NoStop}%
\bibitem [{\citenamefont {Flayac}\ \emph {et~al.}(2015)\citenamefont {Flayac},
  \citenamefont {Gerace},\ and\ \citenamefont {Savona}}]{flayac}%
  \BibitemOpen
  \bibfield  {author} {\bibinfo {author} {\bibfnamefont {H.}~\bibnamefont
  {Flayac}}, \bibinfo {author} {\bibfnamefont {D.}~\bibnamefont {Gerace}}, \
  and\ \bibinfo {author} {\bibfnamefont {V.}~\bibnamefont {Savona}},\
  }\href@noop {} {\bibfield  {journal} {\bibinfo  {journal} {Sci. Rep.}\
  }\textbf {\bibinfo {volume} {5}},\ \bibinfo {pages} {11223} (\bibinfo {year}
  {2015})}\BibitemShut {NoStop}%
\bibitem [{\citenamefont {Lund-Hansen}\ \emph {et~al.}(2008)\citenamefont
  {Lund-Hansen}, \citenamefont {Stobbe}, \citenamefont {Julsgaard},
  \citenamefont {Thyrrestrup}, \citenamefont {S{\"u}nner}, \citenamefont
  {Kamp}, \citenamefont {Forchel},\ and\ \citenamefont {Lodahl}}]{lodahl4}%
  \BibitemOpen
  \bibfield  {author} {\bibinfo {author} {\bibfnamefont {T.}~\bibnamefont
  {Lund-Hansen}}, \bibinfo {author} {\bibfnamefont {S.}~\bibnamefont {Stobbe}},
  \bibinfo {author} {\bibfnamefont {B.}~\bibnamefont {Julsgaard}}, \bibinfo
  {author} {\bibfnamefont {H.}~\bibnamefont {Thyrrestrup}}, \bibinfo {author}
  {\bibfnamefont {T.}~\bibnamefont {S{\"u}nner}}, \bibinfo {author}
  {\bibfnamefont {M.}~\bibnamefont {Kamp}}, \bibinfo {author} {\bibfnamefont
  {A.}~\bibnamefont {Forchel}}, \ and\ \bibinfo {author} {\bibfnamefont
  {P.}~\bibnamefont {Lodahl}},\ }\href@noop {} {\bibfield  {journal} {\bibinfo
  {journal} {Phys. Rev. Lett.}\ }\textbf {\bibinfo {volume} {101}},\ \bibinfo
  {pages} {113903} (\bibinfo {year} {2008})}\BibitemShut {NoStop}%
\bibitem [{\citenamefont {Yao}\ \emph {et~al.}(2010{\natexlab{a}})\citenamefont
  {Yao}, \citenamefont {{V.S.C. Manga Rao}},\ and\ \citenamefont
  {Hughes}}]{hughes3}%
  \BibitemOpen
  \bibfield  {author} {\bibinfo {author} {\bibfnamefont {P.}~\bibnamefont
  {Yao}}, \bibinfo {author} {\bibnamefont {{V.S.C. Manga Rao}}}, \ and\
  \bibinfo {author} {\bibfnamefont {S.}~\bibnamefont {Hughes}},\ }\href@noop {}
  {\bibfield  {journal} {\bibinfo  {journal} {Laser Photon. Rev.}\ }\textbf
  {\bibinfo {volume} {4}},\ \bibinfo {pages} {499} (\bibinfo {year}
  {2010}{\natexlab{a}})}\BibitemShut {NoStop}%
\bibitem [{\citenamefont {Baba}(2008)}]{baba}%
  \BibitemOpen
  \bibfield  {author} {\bibinfo {author} {\bibfnamefont {T.}~\bibnamefont
  {Baba}},\ }\href@noop {} {\bibfield  {journal} {\bibinfo  {journal} {Nat.
  Photon.}\ }\textbf {\bibinfo {volume} {2}},\ \bibinfo {pages} {465} (\bibinfo
  {year} {2008})}\BibitemShut {NoStop}%
\bibitem [{\citenamefont {Krauss}(2008)}]{krauss}%
  \BibitemOpen
  \bibfield  {author} {\bibinfo {author} {\bibfnamefont {T.~F.}\ \bibnamefont
  {Krauss}},\ }\href@noop {} {\bibfield  {journal} {\bibinfo  {journal} {Nat.
  Photon.}\ }\textbf {\bibinfo {volume} {2}},\ \bibinfo {pages} {448} (\bibinfo
  {year} {2008})}\BibitemShut {NoStop}%
\bibitem [{\citenamefont {Colman}\ \emph {et~al.}(2010)\citenamefont {Colman},
  \citenamefont {Husko}, \citenamefont {Combri{\'e}}, \citenamefont {Sagnes},
  \citenamefont {Wong},\ and\ \citenamefont {Rossi}}]{colman}%
  \BibitemOpen
  \bibfield  {author} {\bibinfo {author} {\bibfnamefont {P.}~\bibnamefont
  {Colman}}, \bibinfo {author} {\bibfnamefont {C.}~\bibnamefont {Husko}},
  \bibinfo {author} {\bibfnamefont {S.}~\bibnamefont {Combri{\'e}}}, \bibinfo
  {author} {\bibfnamefont {I.}~\bibnamefont {Sagnes}}, \bibinfo {author}
  {\bibfnamefont {C.~W.}\ \bibnamefont {Wong}}, \ and\ \bibinfo {author}
  {\bibfnamefont {A.~D.}\ \bibnamefont {Rossi}},\ }\href@noop {} {\bibfield
  {journal} {\bibinfo  {journal} {Nat. Photon.}\ }\textbf {\bibinfo {volume}
  {4}},\ \bibinfo {pages} {862} (\bibinfo {year} {2010})}\BibitemShut {NoStop}%
\bibitem [{\citenamefont {Monat}\ \emph {et~al.}(2011)\citenamefont {Monat},
  \citenamefont {Spurny}, \citenamefont {Grillet}, \citenamefont
  {O{\rq}Faolain}, \citenamefont {Krauss}, \citenamefont {Eggleton},
  \citenamefont {Bulla}, \citenamefont {Madden},\ and\ \citenamefont
  {Luther-Davies}}]{monat}%
  \BibitemOpen
  \bibfield  {author} {\bibinfo {author} {\bibfnamefont {C.}~\bibnamefont
  {Monat}}, \bibinfo {author} {\bibfnamefont {M.}~\bibnamefont {Spurny}},
  \bibinfo {author} {\bibfnamefont {C.}~\bibnamefont {Grillet}}, \bibinfo
  {author} {\bibfnamefont {L.}~\bibnamefont {O{\rq}Faolain}}, \bibinfo {author}
  {\bibfnamefont {T.~F.}\ \bibnamefont {Krauss}}, \bibinfo {author}
  {\bibfnamefont {B.~J.}\ \bibnamefont {Eggleton}}, \bibinfo {author}
  {\bibfnamefont {D.}~\bibnamefont {Bulla}}, \bibinfo {author} {\bibfnamefont
  {S.}~\bibnamefont {Madden}}, \ and\ \bibinfo {author} {\bibfnamefont
  {B.}~\bibnamefont {Luther-Davies}},\ }\href@noop {} {\bibfield  {journal}
  {\bibinfo  {journal} {Opt. Lett.}\ }\textbf {\bibinfo {volume} {36}},\
  \bibinfo {pages} {2818} (\bibinfo {year} {2011})}\BibitemShut {NoStop}%
\bibitem [{\citenamefont {Gerace}\ and\ \citenamefont
  {Andreani}(2004{\natexlab{a}})}]{gerace}%
  \BibitemOpen
  \bibfield  {author} {\bibinfo {author} {\bibfnamefont {D.}~\bibnamefont
  {Gerace}}\ and\ \bibinfo {author} {\bibfnamefont {L.~C.}\ \bibnamefont
  {Andreani}},\ }\href@noop {} {\bibfield  {journal} {\bibinfo  {journal} {Opt.
  Lett.}\ }\textbf {\bibinfo {volume} {29}},\ \bibinfo {pages} {1897} (\bibinfo
  {year} {2004}{\natexlab{a}})}\BibitemShut {NoStop}%
\bibitem [{\citenamefont {O{\rq}Faolain}\ \emph {et~al.}(2007)\citenamefont
  {O{\rq}Faolain}, \citenamefont {White}, \citenamefont {O{\rq}Brien},
  \citenamefont {Yuan}, \citenamefont {Settle},\ and\ \citenamefont
  {Krauss}}]{faolain}%
  \BibitemOpen
  \bibfield  {author} {\bibinfo {author} {\bibfnamefont {L.}~\bibnamefont
  {O{\rq}Faolain}}, \bibinfo {author} {\bibfnamefont {T.~P.}\ \bibnamefont
  {White}}, \bibinfo {author} {\bibfnamefont {D.}~\bibnamefont {O{\rq}Brien}},
  \bibinfo {author} {\bibfnamefont {X.}~\bibnamefont {Yuan}}, \bibinfo {author}
  {\bibfnamefont {M.~D.}\ \bibnamefont {Settle}}, \ and\ \bibinfo {author}
  {\bibfnamefont {T.~F.}\ \bibnamefont {Krauss}},\ }\href@noop {} {\bibfield
  {journal} {\bibinfo  {journal} {Opt. Express}\ }\textbf {\bibinfo {volume}
  {15}},\ \bibinfo {pages} {13129} (\bibinfo {year} {2007})}\BibitemShut
  {NoStop}%
\bibitem [{\citenamefont {LeThomas}\ \emph {et~al.}(2009)\citenamefont
  {LeThomas}, \citenamefont {Zhang}, \citenamefont {J{\'a}gersk{\'a}},
  \citenamefont {Zabelin}, \citenamefont {Houdr{\'e}}, \citenamefont {Sagnes},\
  and\ \citenamefont {Talneau}}]{talneau}%
  \BibitemOpen
  \bibfield  {author} {\bibinfo {author} {\bibfnamefont {N.}~\bibnamefont
  {LeThomas}}, \bibinfo {author} {\bibfnamefont {H.}~\bibnamefont {Zhang}},
  \bibinfo {author} {\bibfnamefont {J.}~\bibnamefont {J{\'a}gersk{\'a}}},
  \bibinfo {author} {\bibfnamefont {V.}~\bibnamefont {Zabelin}}, \bibinfo
  {author} {\bibfnamefont {R.}~\bibnamefont {Houdr{\'e}}}, \bibinfo {author}
  {\bibfnamefont {I.}~\bibnamefont {Sagnes}}, \ and\ \bibinfo {author}
  {\bibfnamefont {A.}~\bibnamefont {Talneau}},\ }\href@noop {} {\bibfield
  {journal} {\bibinfo  {journal} {Phys. Rev. B}\ }\textbf {\bibinfo {volume}
  {80}},\ \bibinfo {pages} {125332} (\bibinfo {year} {2009})}\BibitemShut
  {NoStop}%
\bibitem [{\citenamefont {Hughes}\ \emph {et~al.}(2005)\citenamefont {Hughes},
  \citenamefont {Ramunno}, \citenamefont {Young},\ and\ \citenamefont
  {Sipe}}]{hughes4}%
  \BibitemOpen
  \bibfield  {author} {\bibinfo {author} {\bibfnamefont {S.}~\bibnamefont
  {Hughes}}, \bibinfo {author} {\bibfnamefont {L.}~\bibnamefont {Ramunno}},
  \bibinfo {author} {\bibfnamefont {J.~F.}\ \bibnamefont {Young}}, \ and\
  \bibinfo {author} {\bibfnamefont {J.~E.}\ \bibnamefont {Sipe}},\ }\href@noop
  {} {\bibfield  {journal} {\bibinfo  {journal} {Phys. Rev. Lett.}\ }\textbf
  {\bibinfo {volume} {94}},\ \bibinfo {pages} {033903} (\bibinfo {year}
  {2005})}\BibitemShut {NoStop}%
\bibitem [{\citenamefont {Kuramochi}\ \emph {et~al.}(2005)\citenamefont
  {Kuramochi}, \citenamefont {Notomi}, \citenamefont {Hughes}, \citenamefont
  {Shinya}, \citenamefont {Watanabe},\ and\ \citenamefont {Ramunno}}]{ramunno}%
  \BibitemOpen
  \bibfield  {author} {\bibinfo {author} {\bibfnamefont {E.}~\bibnamefont
  {Kuramochi}}, \bibinfo {author} {\bibfnamefont {M.}~\bibnamefont {Notomi}},
  \bibinfo {author} {\bibfnamefont {S.}~\bibnamefont {Hughes}}, \bibinfo
  {author} {\bibfnamefont {A.}~\bibnamefont {Shinya}}, \bibinfo {author}
  {\bibfnamefont {T.}~\bibnamefont {Watanabe}}, \ and\ \bibinfo {author}
  {\bibfnamefont {L.}~\bibnamefont {Ramunno}},\ }\href@noop {} {\bibfield
  {journal} {\bibinfo  {journal} {Phys. Rev. B}\ }\textbf {\bibinfo {volume}
  {72}},\ \bibinfo {pages} {161318(R)} (\bibinfo {year} {2005})}\BibitemShut
  {NoStop}%
\bibitem [{\citenamefont {Mazoyer}\ \emph
  {et~al.}(2009{\natexlab{a}})\citenamefont {Mazoyer}, \citenamefont
  {Hugonin},\ and\ \citenamefont {Lalanne}}]{lalanne1}%
  \BibitemOpen
  \bibfield  {author} {\bibinfo {author} {\bibfnamefont {S.}~\bibnamefont
  {Mazoyer}}, \bibinfo {author} {\bibfnamefont {J.~P.}\ \bibnamefont
  {Hugonin}}, \ and\ \bibinfo {author} {\bibfnamefont {P.}~\bibnamefont
  {Lalanne}},\ }\href@noop {} {\bibfield  {journal} {\bibinfo  {journal} {Phys.
  Rev. Lett.}\ }\textbf {\bibinfo {volume} {103}},\ \bibinfo {pages} {063903}
  (\bibinfo {year} {2009}{\natexlab{a}})}\BibitemShut {NoStop}%
\bibitem [{\citenamefont {Patterson}\ \emph
  {et~al.}(2009{\natexlab{a}})\citenamefont {Patterson}, \citenamefont
  {Hughes}, \citenamefont {Combri{\'e}}, \citenamefont {Tran}, \citenamefont
  {{De Rossi}}, \citenamefont {Gabet},\ and\ \citenamefont
  {Jaou{\"e}n}}]{patterson1}%
  \BibitemOpen
  \bibfield  {author} {\bibinfo {author} {\bibfnamefont {M.}~\bibnamefont
  {Patterson}}, \bibinfo {author} {\bibfnamefont {S.}~\bibnamefont {Hughes}},
  \bibinfo {author} {\bibfnamefont {S.}~\bibnamefont {Combri{\'e}}}, \bibinfo
  {author} {\bibfnamefont {N.-V.-Quynh}\ \bibnamefont {Tran}}, \bibinfo {author}
  {\bibfnamefont {A.}~\bibnamefont {{De Rossi}}}, \bibinfo {author}
  {\bibfnamefont {R.}~\bibnamefont {Gabet}}, \ and\ \bibinfo {author}
  {\bibfnamefont {Y.}~\bibnamefont {Jaou{\"e}n}},\ }\href@noop {} {\bibfield
  {journal} {\bibinfo  {journal} {Phys. Rev. Lett.}\ }\textbf {\bibinfo
  {volume} {102}},\ \bibinfo {pages} {253903} (\bibinfo {year}
  {2009}{\natexlab{a}})}\BibitemShut {NoStop}%
\bibitem [{\citenamefont {Schwartz}\ \emph {et~al.}(2007)\citenamefont
  {Schwartz}, \citenamefont {Bartal}, \citenamefont {Fishman},\ and\
  \citenamefont {Segev}}]{segev}%
  \BibitemOpen
  \bibfield  {author} {\bibinfo {author} {\bibfnamefont {T.}~\bibnamefont
  {Schwartz}}, \bibinfo {author} {\bibfnamefont {G.}~\bibnamefont {Bartal}},
  \bibinfo {author} {\bibfnamefont {S.}~\bibnamefont {Fishman}}, \ and\
  \bibinfo {author} {\bibfnamefont {M.}~\bibnamefont {Segev}},\ }\href@noop {}
  {\bibfield  {journal} {\bibinfo  {journal} {Nature (London)}\ }\textbf
  {\bibinfo {volume} {446}},\ \bibinfo {pages} {52} (\bibinfo {year}
  {2007})}\BibitemShut {NoStop}%
\bibitem [{\citenamefont {Sapienza}\ \emph {et~al.}(2010)\citenamefont
  {Sapienza}, \citenamefont {Thyrrestrup}, \citenamefont {Stobbe},
  \citenamefont {Garcia}, \citenamefont {Smolka},\ and\ \citenamefont
  {Lodahl}}]{sapienza}%
  \BibitemOpen
  \bibfield  {author} {\bibinfo {author} {\bibfnamefont {L.}~\bibnamefont
  {Sapienza}}, \bibinfo {author} {\bibfnamefont {H.}~\bibnamefont
  {Thyrrestrup}}, \bibinfo {author} {\bibfnamefont {S.}~\bibnamefont {Stobbe}},
  \bibinfo {author} {\bibfnamefont {P.~D.}\ \bibnamefont {Garcia}}, \bibinfo
  {author} {\bibfnamefont {S.}~\bibnamefont {Smolka}}, \ and\ \bibinfo {author}
  {\bibfnamefont {P.}~\bibnamefont {Lodahl}},\ }\href@noop {} {\bibfield
  {journal} {\bibinfo  {journal} {Science}\ }\textbf {\bibinfo {volume}
  {327}},\ \bibinfo {pages} {1352} (\bibinfo {year} {2010})}\BibitemShut
  {NoStop}%
\bibitem [{\citenamefont {Liu}\ \emph {et~al.}(2014)\citenamefont {Liu},
  \citenamefont {Garc{\'i}a}, \citenamefont {Ek}, \citenamefont {Gregersen},
  \citenamefont {Suhr}, \citenamefont {Schubert}, \citenamefont {M{\o}rk},
  \citenamefont {Stobbe},\ and\ \citenamefont {Lodahl}}]{liu}%
  \BibitemOpen
  \bibfield  {author} {\bibinfo {author} {\bibfnamefont {J.}~\bibnamefont
  {Liu}}, \bibinfo {author} {\bibfnamefont {P.~D.}\ \bibnamefont {Garc{\'i}a}},
  \bibinfo {author} {\bibfnamefont {S.}~\bibnamefont {Ek}}, \bibinfo {author}
  {\bibfnamefont {N.}~\bibnamefont {Gregersen}}, \bibinfo {author}
  {\bibfnamefont {T.}~\bibnamefont {Suhr}}, \bibinfo {author} {\bibfnamefont
  {M.}~\bibnamefont {Schubert}}, \bibinfo {author} {\bibfnamefont
  {J.}~\bibnamefont {M{\o}rk}}, \bibinfo {author} {\bibfnamefont
  {S.}~\bibnamefont {Stobbe}}, \ and\ \bibinfo {author} {\bibfnamefont
  {P.}~\bibnamefont {Lodahl}},\ }\href@noop {} {\bibfield  {journal} {\bibinfo
  {journal} {Nat. Nan.}\ }\textbf {\bibinfo {volume} {9}},\ \bibinfo {pages}
  {285} (\bibinfo {year} {2014})}\BibitemShut {NoStop}%
\bibitem [{\citenamefont {Garc{\'i}a}\ and\ \citenamefont
  {Lodahl}(2016)}]{garcia}%
  \BibitemOpen
  \bibfield  {author} {\bibinfo {author} {\bibfnamefont {P.~D.}\ \bibnamefont
  {Garc{\'i}a}}\ and\ \bibinfo {author} {\bibfnamefont {P.}~\bibnamefont
  {Lodahl}},\ }\href@noop {} {\bibfield  {journal} {\bibinfo  {journal}
  {arXiv:1611.02038}\ } (\bibinfo {year} {2016})}\BibitemShut {NoStop}%
\bibitem [{\citenamefont {Mann}\ \emph
  {et~al.}(2015{\natexlab{a}})\citenamefont {Mann}, \citenamefont {Javadi},
  \citenamefont {Garc{\'i}a}, \citenamefont {Lodahl},\ and\ \citenamefont
  {Hughes}}]{nishan1}%
  \BibitemOpen
  \bibfield  {author} {\bibinfo {author} {\bibfnamefont {N.}~\bibnamefont
  {Mann}}, \bibinfo {author} {\bibfnamefont {A.}~\bibnamefont {Javadi}},
  \bibinfo {author} {\bibfnamefont {P.~D.}\ \bibnamefont {Garc{\'i}a}},
  \bibinfo {author} {\bibfnamefont {P.}~\bibnamefont {Lodahl}}, \ and\ \bibinfo
  {author} {\bibfnamefont {S.}~\bibnamefont {Hughes}},\ }\href@noop {}
  {\bibfield  {journal} {\bibinfo  {journal} {Phys. Rev. A}\ }\textbf {\bibinfo
  {volume} {92}},\ \bibinfo {pages} {023849} (\bibinfo {year}
  {2015}{\natexlab{a}})}\BibitemShut {NoStop}%
\bibitem [{\citenamefont {Topolancik}\ \emph
  {et~al.}(2007{\natexlab{a}})\citenamefont {Topolancik}, \citenamefont
  {Vollmer},\ and\ \citenamefont {Ilic}}]{topolancik1}%
  \BibitemOpen
  \bibfield  {author} {\bibinfo {author} {\bibfnamefont {J.}~\bibnamefont
  {Topolancik}}, \bibinfo {author} {\bibfnamefont {F.}~\bibnamefont {Vollmer}},
  \ and\ \bibinfo {author} {\bibfnamefont {B.}~\bibnamefont {Ilic}},\
  }\href@noop {} {\bibfield  {journal} {\bibinfo  {journal} {Appl. Phys.
  Lett.}\ }\textbf {\bibinfo {volume} {91}},\ \bibinfo {pages} {201102}
  (\bibinfo {year} {2007}{\natexlab{a}})}\BibitemShut {NoStop}%
\bibitem [{\citenamefont {Topolancik}\ \emph
  {et~al.}(2007{\natexlab{b}})\citenamefont {Topolancik}, \citenamefont
  {Ilic},\ and\ \citenamefont {Vollmer}}]{topolancik2}%
  \BibitemOpen
  \bibfield  {author} {\bibinfo {author} {\bibfnamefont {J.}~\bibnamefont
  {Topolancik}}, \bibinfo {author} {\bibfnamefont {B.}~\bibnamefont {Ilic}}, \
  and\ \bibinfo {author} {\bibfnamefont {F.}~\bibnamefont {Vollmer}},\
  }\href@noop {} {\bibfield  {journal} {\bibinfo  {journal} {Phys. Rev. Lett.}\
  }\textbf {\bibinfo {volume} {99}},\ \bibinfo {pages} {253901} (\bibinfo
  {year} {2007}{\natexlab{b}})}\BibitemShut {NoStop}%
\bibitem [{\citenamefont {Hsieh}\ \emph {et~al.}(2015)\citenamefont {Hsieh},
  \citenamefont {Chung}, \citenamefont {McMillan}, \citenamefont {Tsai},
  \citenamefont {Lu}, \citenamefont {Panoiu},\ and\ \citenamefont
  {Wong}}]{hsieh}%
  \BibitemOpen
  \bibfield  {author} {\bibinfo {author} {\bibfnamefont {P.}~\bibnamefont
  {Hsieh}}, \bibinfo {author} {\bibfnamefont {C.}~\bibnamefont {Chung}},
  \bibinfo {author} {\bibfnamefont {J.~F.}\ \bibnamefont {McMillan}}, \bibinfo
  {author} {\bibfnamefont {M.}~\bibnamefont {Tsai}}, \bibinfo {author}
  {\bibfnamefont {M.}~\bibnamefont {Lu}}, \bibinfo {author} {\bibfnamefont
  {N.~C.}\ \bibnamefont {Panoiu}}, \ and\ \bibinfo {author} {\bibfnamefont
  {C.~W.}\ \bibnamefont {Wong}},\ }\href@noop {} {\bibfield  {journal}
  {\bibinfo  {journal} {Nat. Phys.}\ }\textbf {\bibinfo {volume} {11}},\
  \bibinfo {pages} {268} (\bibinfo {year} {2015})}\BibitemShut {NoStop}%
\bibitem [{\citenamefont {Garc{\'i}a}\ \emph {et~al.}(2013)\citenamefont
  {Garc{\'i}a}, \citenamefont {Javadi}, \citenamefont {Thyrrestrup},\ and\
  \citenamefont {Lodahl}}]{Garcia1}%
  \BibitemOpen
  \bibfield  {author} {\bibinfo {author} {\bibfnamefont {P.~D.}\ \bibnamefont
  {Garc{\'i}a}}, \bibinfo {author} {\bibfnamefont {A.}~\bibnamefont {Javadi}},
  \bibinfo {author} {\bibfnamefont {H.}~\bibnamefont {Thyrrestrup}}, \ and\
  \bibinfo {author} {\bibfnamefont {P.}~\bibnamefont {Lodahl}},\ }\href@noop {}
  {\bibfield  {journal} {\bibinfo  {journal} {Appl. Phys. Lett.}\ }\textbf
  {\bibinfo {volume} {102}},\ \bibinfo {pages} {031101} (\bibinfo {year}
  {2013})}\BibitemShut {NoStop}%
\bibitem [{\citenamefont {Faggiani}\ \emph {et~al.}(2016)\citenamefont
  {Faggiani}, \citenamefont {Baron}, \citenamefont {Zang}, \citenamefont
  {Lalouat}, \citenamefont {Schulz}, \citenamefont {O{\rq}Regan}, \citenamefont
  {Vynck}, \citenamefont {Cluzel}, \citenamefont {de~Fornel}, \citenamefont
  {Krauss},\ and\ \citenamefont {Lalanne}}]{lalanne2}%
  \BibitemOpen
  \bibfield  {author} {\bibinfo {author} {\bibfnamefont {R.}~\bibnamefont
  {Faggiani}}, \bibinfo {author} {\bibfnamefont {A.}~\bibnamefont {Baron}},
  \bibinfo {author} {\bibfnamefont {X.}~\bibnamefont {Zang}}, \bibinfo {author}
  {\bibfnamefont {L.}~\bibnamefont {Lalouat}}, \bibinfo {author} {\bibfnamefont
  {S.~A.}\ \bibnamefont {Schulz}}, \bibinfo {author} {\bibfnamefont
  {B.}~\bibnamefont {O{\rq}Regan}}, \bibinfo {author} {\bibfnamefont
  {K.}~\bibnamefont {Vynck}}, \bibinfo {author} {\bibfnamefont
  {B.}~\bibnamefont {Cluzel}}, \bibinfo {author} {\bibfnamefont
  {F.}~\bibnamefont {de~Fornel}}, \bibinfo {author} {\bibfnamefont {T.~F.}\
  \bibnamefont {Krauss}}, \ and\ \bibinfo {author} {\bibfnamefont
  {P.}~\bibnamefont {Lalanne}},\ }\href@noop {} {\bibfield  {journal} {\bibinfo
   {journal} {Sci. Rep.}\ }\textbf {\bibinfo {volume} {6}},\ \bibinfo {pages}
  {27037} (\bibinfo {year} {2016})}\BibitemShut {NoStop}%
\bibitem [{\citenamefont {Thyrrestrup}\ \emph {et~al.}(2012)\citenamefont
  {Thyrrestrup}, \citenamefont {Smolka}, \citenamefont {Sapienza},\ and\
  \citenamefont {Lodahl}}]{lodahl1}%
  \BibitemOpen
  \bibfield  {author} {\bibinfo {author} {\bibfnamefont {H.}~\bibnamefont
  {Thyrrestrup}}, \bibinfo {author} {\bibfnamefont {S.}~\bibnamefont {Smolka}},
  \bibinfo {author} {\bibfnamefont {L.}~\bibnamefont {Sapienza}}, \ and\
  \bibinfo {author} {\bibfnamefont {P.}~\bibnamefont {Lodahl}},\ }\href@noop {}
  {\bibfield  {journal} {\bibinfo  {journal} {Phys. Rev. Lett.}\ }\textbf
  {\bibinfo {volume} {108}},\ \bibinfo {pages} {113901} (\bibinfo {year}
  {2012})}\BibitemShut {NoStop}%
\bibitem [{\citenamefont {Smolka}\ \emph {et~al.}(2011)\citenamefont {Smolka},
  \citenamefont {Thyrrestrup}, \citenamefont {Sapienza}, \citenamefont
  {Lehmann}, \citenamefont {Rix}, \citenamefont {Froufe-P{\'e}rez},
  \citenamefont {Garc{\'i}a},\ and\ \citenamefont {Lodahl}}]{lodahl2}%
  \BibitemOpen
  \bibfield  {author} {\bibinfo {author} {\bibfnamefont {S.}~\bibnamefont
  {Smolka}}, \bibinfo {author} {\bibfnamefont {H.}~\bibnamefont {Thyrrestrup}},
  \bibinfo {author} {\bibfnamefont {L.}~\bibnamefont {Sapienza}}, \bibinfo
  {author} {\bibfnamefont {T.~B.}\ \bibnamefont {Lehmann}}, \bibinfo {author}
  {\bibfnamefont {K.~R.}\ \bibnamefont {Rix}}, \bibinfo {author} {\bibfnamefont
  {L.~S.}\ \bibnamefont {Froufe-P{\'e}rez}}, \bibinfo {author} {\bibfnamefont
  {P.~D.}\ \bibnamefont {Garc{\'i}a}}, \ and\ \bibinfo {author} {\bibfnamefont
  {P.}~\bibnamefont {Lodahl}},\ }\href@noop {} {\bibfield  {journal} {\bibinfo
  {journal} {New J. Phys.}\ }\textbf {\bibinfo {volume} {13}},\ \bibinfo
  {pages} {063044} (\bibinfo {year} {2011})}\BibitemShut {NoStop}%
\bibitem [{\citenamefont {Javadi}\ \emph {et~al.}(2014)\citenamefont {Javadi},
  \citenamefont {Maibom}, \citenamefont {Sapienza}, \citenamefont
  {Thyrrestrup}, \citenamefont {Garc{\'i}a},\ and\ \citenamefont
  {Lodahl}}]{lodahl3}%
  \BibitemOpen
  \bibfield  {author} {\bibinfo {author} {\bibfnamefont {A.}~\bibnamefont
  {Javadi}}, \bibinfo {author} {\bibfnamefont {S.}~\bibnamefont {Maibom}},
  \bibinfo {author} {\bibfnamefont {L.}~\bibnamefont {Sapienza}}, \bibinfo
  {author} {\bibfnamefont {H.}~\bibnamefont {Thyrrestrup}}, \bibinfo {author}
  {\bibfnamefont {P.~D.}\ \bibnamefont {Garc{\'i}a}}, \ and\ \bibinfo {author}
  {\bibfnamefont {P.}~\bibnamefont {Lodahl}},\ }\href@noop {} {\bibfield
  {journal} {\bibinfo  {journal} {Opt. Express}\ }\textbf {\bibinfo {volume}
  {22}},\ \bibinfo {pages} {30992} (\bibinfo {year} {2014})}\BibitemShut
  {NoStop}%
\bibitem [{\citenamefont {Savona}(2011)}]{savona1}%
  \BibitemOpen
  \bibfield  {author} {\bibinfo {author} {\bibfnamefont {V.}~\bibnamefont
  {Savona}},\ }\href@noop {} {\bibfield  {journal} {\bibinfo  {journal} {Phys.
  Rev. B}\ }\textbf {\bibinfo {volume} {83}},\ \bibinfo {pages} {085301}
  (\bibinfo {year} {2011})}\BibitemShut {NoStop}%
\bibitem [{\citenamefont {Savona}(2012)}]{savona2}%
  \BibitemOpen
  \bibfield  {author} {\bibinfo {author} {\bibfnamefont {V.}~\bibnamefont
  {Savona}},\ }\href@noop {} {\bibfield  {journal} {\bibinfo  {journal} {Phys.
  Rev. B}\ }\textbf {\bibinfo {volume} {86}},\ \bibinfo {pages} {079907(E)}
  (\bibinfo {year} {2012})}\BibitemShut {NoStop}%
\bibitem [{\citenamefont {Minkov}\ and\ \citenamefont
  {Savona}(2013{\natexlab{a}})}]{minkov1}%
  \BibitemOpen
  \bibfield  {author} {\bibinfo {author} {\bibfnamefont {M.}~\bibnamefont
  {Minkov}}\ and\ \bibinfo {author} {\bibfnamefont {V.}~\bibnamefont
  {Savona}},\ }\href@noop {} {\bibfield  {journal} {\bibinfo  {journal} {Phys.
  Rev. B}\ }\textbf {\bibinfo {volume} {88}},\ \bibinfo {pages} {081303(R)}
  (\bibinfo {year} {2013}{\natexlab{a}})}\BibitemShut {NoStop}%
\bibitem [{\citenamefont {Andreani}\ and\ \citenamefont
  {Gerace}(2006)}]{andreani}%
  \BibitemOpen
  \bibfield  {author} {\bibinfo {author} {\bibfnamefont {L.~C.}\ \bibnamefont
  {Andreani}}\ and\ \bibinfo {author} {\bibfnamefont {D.}~\bibnamefont
  {Gerace}},\ }\href@noop {} {\bibfield  {journal} {\bibinfo  {journal} {Phys.
  Rev. B}\ }\textbf {\bibinfo {volume} {73}},\ \bibinfo {pages} {235114}
  (\bibinfo {year} {2006})}\BibitemShut {NoStop}%
\bibitem [{\citenamefont {Andreani}\ and\ \citenamefont {Agio}(2003)}]{agio}%
  \BibitemOpen
  \bibfield  {author} {\bibinfo {author} {\bibfnamefont {L.~C.}\ \bibnamefont
  {Andreani}}\ and\ \bibinfo {author} {\bibfnamefont {M.}~\bibnamefont
  {Agio}},\ }\href@noop {} {\bibfield  {journal} {\bibinfo  {journal} {Appl.
  Phys. Lett.}\ }\textbf {\bibinfo {volume} {82}},\ \bibinfo {pages} {2011}
  (\bibinfo {year} {2003})}\BibitemShut {NoStop}%
\bibitem [{\citenamefont {Kristensen}\ and\ \citenamefont
  {Hughes}(2014)}]{kristensen1}%
  \BibitemOpen
  \bibfield  {author} {\bibinfo {author} {\bibfnamefont {P.~T.}\ \bibnamefont
  {Kristensen}}\ and\ \bibinfo {author} {\bibfnamefont {S.}~\bibnamefont
  {Hughes}},\ }\href@noop {} {\bibfield  {journal} {\bibinfo  {journal} {ACS
  Photon.}\ }\textbf {\bibinfo {volume} {1}},\ \bibinfo {pages} {2} (\bibinfo
  {year} {2014})}\BibitemShut {NoStop}%
\bibitem [{\citenamefont {Kristensen}\ \emph {et~al.}(2012)\citenamefont
  {Kristensen}, \citenamefont {Vlack},\ and\ \citenamefont
  {Hughes}}]{kristensen2}%
  \BibitemOpen
  \bibfield  {author} {\bibinfo {author} {\bibfnamefont {P.~T.}\ \bibnamefont
  {Kristensen}}, \bibinfo {author} {\bibfnamefont {C.~V.}\ \bibnamefont
  {Vlack}}, \ and\ \bibinfo {author} {\bibfnamefont {S.}~\bibnamefont
  {Hughes}},\ }\href@noop {} {\bibfield  {journal} {\bibinfo  {journal} {Opt.
  Lett.}\ }\textbf {\bibinfo {volume} {37}},\ \bibinfo {pages} {1649} (\bibinfo
  {year} {2012})}\BibitemShut {NoStop}%
\bibitem [{\citenamefont {Gerace}\ and\ \citenamefont
  {Andreani}(2004{\natexlab{b}})}]{dario1}%
  \BibitemOpen
  \bibfield  {author} {\bibinfo {author} {\bibfnamefont {D.}~\bibnamefont
  {Gerace}}\ and\ \bibinfo {author} {\bibfnamefont {L.~C.}\ \bibnamefont
  {Andreani}},\ }\href@noop {} {\bibfield  {journal} {\bibinfo  {journal} {Opt.
  Lett.}\ }\textbf {\bibinfo {volume} {29}},\ \bibinfo {pages} {1897} (\bibinfo
  {year} {2004}{\natexlab{b}})}\BibitemShut {NoStop}%
\bibitem [{\citenamefont {Mann}\ \emph
  {et~al.}(2015{\natexlab{b}})\citenamefont {Mann}, \citenamefont {Patterson},\
  and\ \citenamefont {Hughes}}]{nishan2}%
  \BibitemOpen
  \bibfield  {author} {\bibinfo {author} {\bibfnamefont {N.}~\bibnamefont
  {Mann}}, \bibinfo {author} {\bibfnamefont {M.}~\bibnamefont {Patterson}}, \
  and\ \bibinfo {author} {\bibfnamefont {S.}~\bibnamefont {Hughes}},\
  }\href@noop {} {\bibfield  {journal} {\bibinfo  {journal} {Phys. Rev. B}\
  }\textbf {\bibinfo {volume} {91}},\ \bibinfo {pages} {245151} (\bibinfo
  {year} {2015}{\natexlab{b}})}\BibitemShut {NoStop}%
\bibitem [{\citenamefont {Minkov}\ \emph {et~al.}(2013)\citenamefont {Minkov},
  \citenamefont {Dharanipathy}, \citenamefont {Houdr{\'e}},\ and\ \citenamefont
  {Savona}}]{minkov2}%
  \BibitemOpen
  \bibfield  {author} {\bibinfo {author} {\bibfnamefont {M.}~\bibnamefont
  {Minkov}}, \bibinfo {author} {\bibfnamefont {U.~P.}\ \bibnamefont
  {Dharanipathy}}, \bibinfo {author} {\bibfnamefont {R.}~\bibnamefont
  {Houdr{\'e}}}, \ and\ \bibinfo {author} {\bibfnamefont {V.}~\bibnamefont
  {Savona}},\ }\href@noop {} {\bibfield  {journal} {\bibinfo  {journal} {Opt.
  Express}\ }\textbf {\bibinfo {volume} {21}},\ \bibinfo {pages} {28233}
  (\bibinfo {year} {2013})}\BibitemShut {NoStop}%
\bibitem [{\citenamefont {Mazoyer}\ \emph
  {et~al.}(2009{\natexlab{b}})\citenamefont {Mazoyer}, \citenamefont
  {Hugonin},\ and\ \citenamefont {Lalanne}}]{mazoyer}%
  \BibitemOpen
  \bibfield  {author} {\bibinfo {author} {\bibfnamefont {S.}~\bibnamefont
  {Mazoyer}}, \bibinfo {author} {\bibfnamefont {J.~P.}\ \bibnamefont
  {Hugonin}}, \ and\ \bibinfo {author} {\bibfnamefont {P.}~\bibnamefont
  {Lalanne}},\ }\href@noop {} {\bibfield  {journal} {\bibinfo  {journal} {Phys.
  Rev. Lett.}\ }\textbf {\bibinfo {volume} {103}},\ \bibinfo {pages} {063903}
  (\bibinfo {year} {2009}{\natexlab{b}})}\BibitemShut {NoStop}%
\bibitem [{\citenamefont {Yao}\ \emph {et~al.}(2010{\natexlab{b}})\citenamefont
  {Yao}, \citenamefont {{V.S.C. Manga Rao}},\ and\ \citenamefont
  {Hughes}}]{yao}%
  \BibitemOpen
  \bibfield  {author} {\bibinfo {author} {\bibfnamefont {P.}~\bibnamefont
  {Yao}}, \bibinfo {author} {\bibnamefont {{V.S.C. Manga Rao}}}, \ and\
  \bibinfo {author} {\bibfnamefont {S.}~\bibnamefont {Hughes}},\ }\href@noop {}
  {\bibfield  {journal} {\bibinfo  {journal} {Laser Photon. Rev.}\ }\textbf
  {\bibinfo {volume} {4}},\ \bibinfo {pages} {499} (\bibinfo {year}
  {2010}{\natexlab{b}})}\BibitemShut {NoStop}%
\bibitem [{\citenamefont {Novotny}\ and\ \citenamefont
  {Hecht}(2006)}]{novotny}%
  \BibitemOpen
  \bibfield  {author} {\bibinfo {author} {\bibfnamefont {L.}~\bibnamefont
  {Novotny}}\ and\ \bibinfo {author} {\bibfnamefont {B.}~\bibnamefont
  {Hecht}},\ }\href@noop {} {\emph {\bibinfo {title} {{Principles of
  nano-optics}}}}\ (\bibinfo  {publisher} {Cambridge University Press},\
  \bibinfo {year} {2006})\BibitemShut {NoStop}%
\bibitem [{\citenamefont {Patterson}\ \emph
  {et~al.}(2009{\natexlab{b}})\citenamefont {Patterson}, \citenamefont
  {Hughes}, \citenamefont {Dalacu},\ and\ \citenamefont
  {Williams}}]{patterson2}%
  \BibitemOpen
  \bibfield  {author} {\bibinfo {author} {\bibfnamefont {M.}~\bibnamefont
  {Patterson}}, \bibinfo {author} {\bibfnamefont {S.}~\bibnamefont {Hughes}},
  \bibinfo {author} {\bibfnamefont {D.}~\bibnamefont {Dalacu}}, \ and\ \bibinfo
  {author} {\bibfnamefont {R.~L.}\ \bibnamefont {Williams}},\ }\href@noop {}
  {\bibfield  {journal} {\bibinfo  {journal} {Phys. Rev. B}\ }\textbf {\bibinfo
  {volume} {80}},\ \bibinfo {pages} {125307} (\bibinfo {year}
  {2009}{\natexlab{b}})}\BibitemShut {NoStop}%
\bibitem [{\citenamefont {Gao}\ \emph {et~al.}(2013)\citenamefont {Gao},
  \citenamefont {Combrie}, \citenamefont {Liang}, \citenamefont
  {Schmitteckert}, \citenamefont {Lehoucq}, \citenamefont {Xavier},
  \citenamefont {Xu}, \citenamefont {Busch}, \citenamefont {Huffaker},
  \citenamefont {Rossi},\ and\ \citenamefont {Wong}}]{wong}%
  \BibitemOpen
  \bibfield  {author} {\bibinfo {author} {\bibfnamefont {J.}~\bibnamefont
  {Gao}}, \bibinfo {author} {\bibfnamefont {S.}~\bibnamefont {Combrie}},
  \bibinfo {author} {\bibfnamefont {B.}~\bibnamefont {Liang}}, \bibinfo
  {author} {\bibfnamefont {P.}~\bibnamefont {Schmitteckert}}, \bibinfo {author}
  {\bibfnamefont {G.}~\bibnamefont {Lehoucq}}, \bibinfo {author} {\bibfnamefont
  {S.}~\bibnamefont {Xavier}}, \bibinfo {author} {\bibfnamefont
  {X.}~\bibnamefont {Xu}}, \bibinfo {author} {\bibfnamefont {K.}~\bibnamefont
  {Busch}}, \bibinfo {author} {\bibfnamefont {D.~L.}\ \bibnamefont {Huffaker}},
  \bibinfo {author} {\bibfnamefont {A.~D.}\ \bibnamefont {Rossi}}, \ and\
  \bibinfo {author} {\bibfnamefont {C.~W.}\ \bibnamefont {Wong}},\ }\href@noop
  {} {\bibfield  {journal} {\bibinfo  {journal} {Sci. Rep.}\ }\textbf {\bibinfo
  {volume} {3}},\ \bibinfo {pages} {1994} (\bibinfo {year} {2013})}\BibitemShut
  {NoStop}%
\bibitem [{\citenamefont {Caz{\'e}}\ \emph {et~al.}(2013)\citenamefont
  {Caz{\'e}}, \citenamefont {Pierrat},\ and\ \citenamefont
  {Carminati}}]{carminati}%
  \BibitemOpen
  \bibfield  {author} {\bibinfo {author} {\bibfnamefont {A.}~\bibnamefont
  {Caz{\'e}}}, \bibinfo {author} {\bibfnamefont {R.}~\bibnamefont {Pierrat}}, \
  and\ \bibinfo {author} {\bibfnamefont {R.}~\bibnamefont {Carminati}},\
  }\href@noop {} {\bibfield  {journal} {\bibinfo  {journal} {Phys. Rev. Lett.}\
  }\textbf {\bibinfo {volume} {111}},\ \bibinfo {pages} {053901} (\bibinfo
  {year} {2013})}\BibitemShut {NoStop}%
\bibitem [{\citenamefont {Kristensen}\ \emph {et~al.}(2011)\citenamefont
  {Kristensen}, \citenamefont {M{\o}rk}, \citenamefont {Lodahl},\ and\
  \citenamefont {Hughes}}]{hughes1}%
  \BibitemOpen
  \bibfield  {author} {\bibinfo {author} {\bibfnamefont {P.~T.}\ \bibnamefont
  {Kristensen}}, \bibinfo {author} {\bibfnamefont {J.}~\bibnamefont {M{\o}rk}},
  \bibinfo {author} {\bibfnamefont {P.}~\bibnamefont {Lodahl}}, \ and\ \bibinfo
  {author} {\bibfnamefont {S.}~\bibnamefont {Hughes}},\ }\href@noop {}
  {\bibfield  {journal} {\bibinfo  {journal} {Phys. Rev. B}\ }\textbf {\bibinfo
  {volume} {83}},\ \bibinfo {pages} {075305} (\bibinfo {year}
  {2011})}\BibitemShut {NoStop}%
\bibitem [{\citenamefont {Stievater}\ \emph {et~al.}(2001)\citenamefont
  {Stievater}, \citenamefont {Li}, \citenamefont {Steel}, \citenamefont
  {Gammon}, \citenamefont {Katzer}, \citenamefont {Park}, \citenamefont
  {Piermarocchi},\ and\ \citenamefont {Sham}}]{stievater}%
  \BibitemOpen
  \bibfield  {author} {\bibinfo {author} {\bibfnamefont {T.~H.}\ \bibnamefont
  {Stievater}}, \bibinfo {author} {\bibfnamefont {X.}~\bibnamefont {Li}},
  \bibinfo {author} {\bibfnamefont {D.~G.}\ \bibnamefont {Steel}}, \bibinfo
  {author} {\bibfnamefont {D.}~\bibnamefont {Gammon}}, \bibinfo {author}
  {\bibfnamefont {D.~S.}\ \bibnamefont {Katzer}}, \bibinfo {author}
  {\bibfnamefont {D.}~\bibnamefont {Park}}, \bibinfo {author} {\bibfnamefont
  {C.}~\bibnamefont {Piermarocchi}}, \ and\ \bibinfo {author} {\bibfnamefont
  {L.~J.}\ \bibnamefont {Sham}},\ }\href@noop {} {\bibfield  {journal}
  {\bibinfo  {journal} {Phys. Rev. Lett.}\ }\textbf {\bibinfo {volume} {87}},\
  \bibinfo {pages} {133603} (\bibinfo {year} {2001})}\BibitemShut {NoStop}%
\bibitem [{\citenamefont {Guest}\ \emph {et~al.}(2002)\citenamefont {Guest},
  \citenamefont {Stievater}, \citenamefont {Li}, \citenamefont {Cheng},
  \citenamefont {Steel}, \citenamefont {Gammon}, \citenamefont {Katzer},
  \citenamefont {Park}, \citenamefont {Ell}, \citenamefont {Thr{\"a}nhardt},
  \citenamefont {Khitrova},\ and\ \citenamefont {Gibbs}}]{jrguest}%
  \BibitemOpen
  \bibfield  {author} {\bibinfo {author} {\bibfnamefont {J.~R.}\ \bibnamefont
  {Guest}}, \bibinfo {author} {\bibfnamefont {T.~H.}\ \bibnamefont
  {Stievater}}, \bibinfo {author} {\bibfnamefont {X.}~\bibnamefont {Li}},
  \bibinfo {author} {\bibfnamefont {J.}~\bibnamefont {Cheng}}, \bibinfo
  {author} {\bibfnamefont {D.~G.}\ \bibnamefont {Steel}}, \bibinfo {author}
  {\bibfnamefont {D.}~\bibnamefont {Gammon}}, \bibinfo {author} {\bibfnamefont
  {D.~S.}\ \bibnamefont {Katzer}}, \bibinfo {author} {\bibfnamefont
  {D.}~\bibnamefont {Park}}, \bibinfo {author} {\bibfnamefont {C.}~\bibnamefont
  {Ell}}, \bibinfo {author} {\bibfnamefont {A.}~\bibnamefont {Thr{\"a}nhardt}},
  \bibinfo {author} {\bibfnamefont {G.}~\bibnamefont {Khitrova}}, \ and\
  \bibinfo {author} {\bibfnamefont {H.~M.}\ \bibnamefont {Gibbs}},\ }\href@noop
  {} {\bibfield  {journal} {\bibinfo  {journal} {Phys. Rev. B}\ }\textbf
  {\bibinfo {volume} {65}},\ \bibinfo {pages} {241310(R)} (\bibinfo {year}
  {2002})}\BibitemShut {NoStop}%
\bibitem [{\citenamefont {Minkov}\ and\ \citenamefont
  {Savona}(2013{\natexlab{b}})}]{minkov4}%
  \BibitemOpen
  \bibfield  {author} {\bibinfo {author} {\bibfnamefont {M.}~\bibnamefont
  {Minkov}}\ and\ \bibinfo {author} {\bibfnamefont {V.}~\bibnamefont
  {Savona}},\ }\href@noop {} {\bibfield  {journal} {\bibinfo  {journal} {Phys.
  Rev. B}\ }\textbf {\bibinfo {volume} {87}},\ \bibinfo {pages} {125306}
  (\bibinfo {year} {2013}{\natexlab{b}})}\BibitemShut {NoStop}%
\bibitem [{\citenamefont {Reithmaier}\ \emph {et~al.}(2004)\citenamefont
  {Reithmaier}, \citenamefont {Sek}, \citenamefont {L{\"o}ffler}, \citenamefont
  {Hofmann}, \citenamefont {Kuhn}, \citenamefont {Reitzenstein}, \citenamefont
  {Keldysh}, \citenamefont {Kulakovskii}, \citenamefont {Reinecke},\ and\
  \citenamefont {Forchel}}]{reithmaier}%
  \BibitemOpen
  \bibfield  {author} {\bibinfo {author} {\bibfnamefont {J.~P.}\ \bibnamefont
  {Reithmaier}}, \bibinfo {author} {\bibfnamefont {G.}~\bibnamefont {Sek}},
  \bibinfo {author} {\bibfnamefont {A.}~\bibnamefont {L{\"o}ffler}}, \bibinfo
  {author} {\bibfnamefont {C.}~\bibnamefont {Hofmann}}, \bibinfo {author}
  {\bibfnamefont {S.}~\bibnamefont {Kuhn}}, \bibinfo {author} {\bibfnamefont
  {S.}~\bibnamefont {Reitzenstein}}, \bibinfo {author} {\bibfnamefont {L.~V.}\
  \bibnamefont {Keldysh}}, \bibinfo {author} {\bibfnamefont {V.~D.}\
  \bibnamefont {Kulakovskii}}, \bibinfo {author} {\bibfnamefont {T.~L.}\
  \bibnamefont {Reinecke}}, \ and\ \bibinfo {author} {\bibfnamefont
  {A.}~\bibnamefont {Forchel}},\ }\href@noop {} {\bibfield  {journal} {\bibinfo
   {journal} {Nature (London)}\ }\textbf {\bibinfo {volume} {432}},\ \bibinfo
  {pages} {197} (\bibinfo {year} {2004})}\BibitemShut {NoStop}%
\bibitem [{\citenamefont {Andreani}\ \emph {et~al.}(1999)\citenamefont
  {Andreani}, \citenamefont {Panzarini},\ and\ \citenamefont
  {G{\'e}rard}}]{andreanisc}%
  \BibitemOpen
  \bibfield  {author} {\bibinfo {author} {\bibfnamefont {L.~C.}\ \bibnamefont
  {Andreani}}, \bibinfo {author} {\bibfnamefont {G.}~\bibnamefont {Panzarini}},
  \ and\ \bibinfo {author} {\bibfnamefont {J.-M.}\ \bibnamefont {G{\'e}rard}},\
  }\href@noop {} {\bibfield  {journal} {\bibinfo  {journal} {Phys. Rev. B}\
  }\textbf {\bibinfo {volume} {60}},\ \bibinfo {pages} {13276} (\bibinfo {year}
  {1999})}\BibitemShut {NoStop}%
\bibitem [{\citenamefont {Hennessy}\ \emph
  {et~al.}(2007{\natexlab{b}})\citenamefont {Hennessy}, \citenamefont
  {Badolato}, \citenamefont {Winger}, \citenamefont {Gerace}, \citenamefont
  {Atat{\"u}re}, \citenamefont {Gulde}, \citenamefont {F{\"a}lt}, \citenamefont
  {Hu},\ and\ \citenamefont {Imamo\v{g}lu}}]{kevin07nat}%
  \BibitemOpen
  \bibfield  {author} {\bibinfo {author} {\bibfnamefont {K.}~\bibnamefont
  {Hennessy}}, \bibinfo {author} {\bibfnamefont {A.}~\bibnamefont {Badolato}},
  \bibinfo {author} {\bibfnamefont {M.}~\bibnamefont {Winger}}, \bibinfo
  {author} {\bibfnamefont {D.}~\bibnamefont {Gerace}}, \bibinfo {author}
  {\bibfnamefont {M.}~\bibnamefont {Atat{\"u}re}}, \bibinfo {author}
  {\bibfnamefont {S.}~\bibnamefont {Gulde}}, \bibinfo {author} {\bibfnamefont
  {S.}~\bibnamefont {F{\"a}lt}}, \bibinfo {author} {\bibfnamefont
  {E.}~\bibnamefont {Hu}}, \ and\ \bibinfo {author} {\bibfnamefont
  {A.}~\bibnamefont {Imamo\v{g}lu}},\ }\href@noop {} {\bibfield  {journal}
  {\bibinfo  {journal} {Nature (London)}\ }\textbf {\bibinfo {volume} {445}},\
  \bibinfo {pages} {896} (\bibinfo {year} {2007}{\natexlab{b}})}\BibitemShut
  {NoStop}%
\bibitem [{\citenamefont {Faraon}\ \emph {et~al.}(2010)\citenamefont {Faraon},
  \citenamefont {Majumdar}, \citenamefont {Kim}, \citenamefont {Petroff},\ and\
  \citenamefont {{Vu\ifmmode \check{c}\else \v{c}\fi{}kovi\ifmmode
  \acute{c}\else {\'c}\fi{}}}}]{faraon}%
  \BibitemOpen
  \bibfield  {author} {\bibinfo {author} {\bibfnamefont {A.}~\bibnamefont
  {Faraon}}, \bibinfo {author} {\bibfnamefont {A.}~\bibnamefont {Majumdar}},
  \bibinfo {author} {\bibfnamefont {H.}~\bibnamefont {Kim}}, \bibinfo {author}
  {\bibfnamefont {P.}~\bibnamefont {Petroff}}, \ and\ \bibinfo {author}
  {\bibfnamefont {J.}~\bibnamefont {{Vu\ifmmode \check{c}\else
  \v{c}\fi{}kovi\ifmmode \acute{c}\else {\'c}\fi{}}}},\ }\href@noop {}
  {\bibfield  {journal} {\bibinfo  {journal} {Phys. Rev. Lett.}\ }\textbf
  {\bibinfo {volume} {104}},\ \bibinfo {pages} {047402} (\bibinfo {year}
  {2010})}\BibitemShut {NoStop}%
\end{thebibliography}
\end{document}